\documentclass[11pt]{article}
\usepackage{amsmath,amssymb}
\usepackage{multirow}
\usepackage{changepage}

\usepackage{textcomp,marvosym}
\usepackage{cite}
\usepackage[numbers,round,semicolon,sort&compress]{natbib}

\usepackage{nameref,hyperref}
\usepackage{float}

\usepackage[right]{lineno}

\usepackage[nopatch=eqnum]{microtype}
\DisableLigatures[f]{encoding = *, family = * }

\usepackage[table]{xcolor}

\usepackage{array}
\usepackage{enumitem}

\newcolumntype{+}{!{\vrule width 2pt}}

\newlength\savedwidth

\raggedright
\usepackage[aboveskip=1pt,labelfont=bf,labelsep=period,justification=raggedright,singlelinecheck=off]{caption}

\makeatletter
\renewcommand{\@biblabel}[1]{\quad#1.}
\makeatother

\usepackage{xstring}

\usepackage{lastpage,fancyhdr,graphicx}
\usepackage{epstopdf}
\usepackage{cleveref}
\fancyheadoffset[L]{2.25in}
\usepackage{booktabs}%

\newlist{todolist}{itemize}{2}
\setlist[todolist]{label=$\square$}

\usepackage[margin=3cm,footskip=2cm]{geometry}
\usepackage{caption}
\usepackage{subcaption}
\usepackage{graphicx}%
\usepackage{multirow}%
\usepackage{amsmath,amssymb,amsfonts}%
\usepackage{amsthm}%
\usepackage{mathrsfs}%
\usepackage{latexsym}%
\usepackage{mathtools}%
\usepackage[title]{appendix}%
\usepackage{xcolor}%
\usepackage{color,cancel}
\usepackage{textcomp}%
\usepackage{manyfoot}%
\usepackage[utf8]{inputenc}
\usepackage{listings}%

\usepackage{tocloft}
\usepackage{authblk}
\usepackage{lipsum}
\usepackage{wrapfig}
\usepackage{enumerate}
\usepackage{soul}
\usepackage{chemarrow}

\usepackage{booktabs}%
\usepackage{multicol}
\usepackage{multirow}
\usepackage[math]{cellspace}
\usepackage{algorithm}%
\usepackage{algorithmicx}%
\usepackage{algpseudocode}%
\usepackage{listings}%
\usepackage{units}
\usepackage{verbatim}
\usepackage{fancyvrb}

\fvset{fontsize=\normalsize}

\usepackage[T1]{fontenc}
\usepackage[sc]{mathpazo}
\setkeys{Gin}{width=\linewidth,totalheight=\textheight,keepaspectratio}

\usepackage{hyperref}
\usepackage{cleveref}

\hypersetup{
	hidelinks = true
}

\usepackage{xstring}
\makeatletter
\newcommand{\getnamereftext}[1]{%
  \@ifundefined{r@#1}{}{%
    \unexpanded\expandafter\expandafter\expandafter{%
      \expandafter\expandafter\expandafter\@thirdoffive\csname r@#1\endcsname
    }%
  }%
}

\newcommand{\shortRef}[1]{%
  \StrBefore{\getnamereftext{#1}}{.}[\myref]%
  \IfBeginWith{\myref}{``}{\StrBetween{\myref}{``}{''}[\myref]}{}%
  \hyperref[#1]{\myref}%
}

\title{Platelet plug microstructure and flow modulate fibrin gelation dynamics: Insights from computational simulations}

\author[1]{Janneke M.H. Cruts}
\author[1,2]{Frank J.H. Gijsen}
\author[3,4]{Aaron L. Fogelson}
\author[5]{Anna C. Nelson \thanks{\url{annanelson@unm.edu}}}

\affil[1]{Department of Biomedical Engineering, Erasmus Medical Center, Rotterdam, the Netherlands}

\affil[2]{Department of Biomechanical Engineering, Delft University of Technology, Delft, the Netherlands}

\affil[3]{Department of Mathematics, University of Utah, Salt Lake City, Utah, United States}

\affil[4]{Department of Biomedical Engineering, University of Utah, Salt Lake City, Utah, United States}

\affil[5]{Department of Mathematics and Statistics, University of New Mexico, Albuquerque, New Mexico, United States}

\begin{document}

\maketitle

\begin{abstract}
During the formation of a thrombus, the architecture of the growing platelet aggregate is heterogeneous, with areas of dense and loosely packed platelets. The surface of activated platelets facilitate biochemical coagulation reactions that ultimately result in the formation of a fibrin network which stabilizes the thrombus. How platelet-plug microstructure and flow jointly govern the onset and development of fibrin is incompletely understood.  We developed a novel 2D computational framework that integrates (1) a pre-adhered, discrete platelet aggregate, (2) a reduced coagulation model that generates thrombin, and (3) a fibrin polymerization model. Three platelet-plug configurations were constructed with prescribed interplatelet gaps and simulations were performed for each with various wall shear rates. We quantified spatiotemporal clotting metrics, including coagulation factor concentrations, fibrin evolution, and gelation onset. Across geometries, gelation initiation accelerated with increasing plug density. For more dense geometries, gelation emerged first near the plug periphery. As the platelet density increased, intraplug transport was increasingly restricted and the thrombin concentrations in the spaces between platelets increased. In contrast, the loose plug supported fibrinogen replenishment deeper into the plug core. Thus, despite slower coagulation initiation due to reduced platelet surface area, monomer generation persisted further into the interior, causing gelation to begin at the vessel wall instead of the periphery. While the proposed model is a simplified representation of blood clot formation under flow, this framework provides a basis for systematic studies of platelet–coagulation interactions under flow, including extensions toward therapeutic developments relevant to prevention of cardiovascular disease.
\end{abstract}

\section*{Author summary}
We present a mathematical model that incorporates three processes which comprise the early stages of blood clot formation: the generation of the enzyme thrombin through coagulation reactions, the formation of a stabilizing fibrin gel, and a preformed platelet aggregate which facilitates coagulation reactions on discrete surfaces. While previous mathematical models have been developed to understand each of these three processes individually, few have investigated them simultaneously while incorporating platelet structure heterogeneity and varying flow conditions. Our results suggest that early fibrin gelation dynamics are influenced heavily by both flow and intraplug architecture, which each affect transport within the aggregate. By treating platelets as discrete surfaces rather than a continuum, we find that chemical accumulation and transport within the preformed platelet aggregate determines the location of the initial fibrin gel. Our simulations show that a dense clot results in a fibrin gel that first appears on the outer edge of the platelet aggregate, with thrombin trapped within the platelet plug interior.  For the loose plug the fibrin gel first appeared close to the injured wall. This mathematical framework provides a valuable, systematic tool to help distinguish effects from mechanisms in blood clot formation and invites future study when higher resolution experimental data becomes available.  
 
\vspace{0.25cm}
\begin{footnotesize}
\noindent \textbf{Keywords:} blood clot formation, thrombus, fibrin polymerization, platelets, mathematical modeling
\end{footnotesize}

\section*{Introduction}\label{sec:intro}
In the event of an injury to the blood vessel wall, exposed material in the subendothelium (SE) triggers the formation of a blood clot, or thrombus, which grows and covers the injury to prevent further bleeding. This hemostatic response to injury involves important biophysical and biochemical processes which includes platelet aggregation and coagulation \cite{furie2008mechanisms,hoffman2001cell}. Platelets are small, disc-shaped cellular components of blood which become activated when they contact collagen exposed on the injured vessel wall.  These platelets can adhere to the vessel wall and release soluble platelet agonists that activate nearby platelets allowing them to cohere to the wall-adherent platelets and to form a platelet plug. The surface of activated platelets supports essential biochemical reactions of the blood coagulation network, which is a large system of enzymatic reactions that occur on the injured vessel wall, the surfaces of activated platelets, and in the fluid near these platelets. Coagulation involves many tightly regulated positive and negative feedback loops with various inhibitors to localize enzyme production, and results in the production of the enzyme thrombin. Thrombin has many functions, including activation of platelets and conversion of the blood-soluble protein fibrinogen into insoluble fibrin monomers that polymerize to form a space-filling, fibrin mesh that stabilizes the growing platelet plug.

The intertwined processes of platelet aggregation, coagulation, and fibrin polymerization occur under a wide range of flow conditions, with shear rates ranging from $<200$ s$^{-1}$ to up to $20,000$ s$^{-1}$ in venous and arterial environments, respectively \cite{fogelson2015fluid}. The development and structure of the growing thrombus can strongly affect the surrounding flow field, and consequently, affect the transport of proteins to, from and within the growing thrombus. These, in turn, impact clot formation, determining how quickly a clot forms and whether it remains intact. Experimental evidence indicates that a growing thrombus is heterogeneous, with a dense inner core of activated platelets able to facilitate coagulation reactions and a less dense shell of aggregated platelets and fibrin polymer \cite{stalker2013hierarchical,welsh2014systems,welsh2016systems,stalker2014systems,tomaiuolo2026volume}. Within this microstructured environment, the conventional understanding of coagulation is that thrombin and subsequently fibrin are generated through both the extrinsic pathway, initiated by tissue factor (TF) exposure, and the intrinsic pathway of coagulation. These are essential for stabilizing the growing thrombus \cite{mackman2007role}. Platelet-mediated coagulation and fibrin generation typically begin adjacent to the vessel wall, near the TF source, and subsequently propagate outward through and beyond the platelet plug \cite{swieringa2016platelet,falati_real-time_2002,furie2008mechanisms}. Additionally, there exists threshold behavior for thrombus initiation, characterized by minimal clot formation at low TF levels and rapid thrombin generation above a critical TF concentration \cite{kuharsky2001surface,diamond2010tissue}. However, the spatial pattern of fibrin formation can be influenced by several factors, including the composition of the injured vessel surface, the local flow conditions (shear rate), and the evolving platelet microstructure \cite{swieringa2016platelet}. The dense packing of activated platelets creates a physical barrier, resulting in restricted access for plasma proteins to the core of this platelet plug \cite{stalker2013hierarchical}. Furthermore, as the thrombus matures, a fibrin cap may form on its surface, imposing an additional transport barrier. This cap spatially separates the procoagulant platelet core from the flowing blood, which contains inactive coagulation factors and platelets \cite{kamocka2010two}. The size of this fibrin cap has been shown to increase as the local shear rate decreases \cite{swieringa2016platelet}. Despite these observations, the exact nature of the interaction between the evolving platelet plug microstructure and the shear rate in affecting the spatiotemporal development of fibrin remains an area requiring further investigation. Furthermore, due to limited resolution of in vitro and in vivo experiments \cite{eyisoylu2026arterial,tomaiuolo2026volume}, it is impossible to precisely quantify how the gap size within a given platelet aggregate impacts the concentration of coagulation and fibrin species at that spatial location. 

Many spatiotemporal computational and mathematical models have been proposed to study aspects of thrombus formation with or without fluid transport \cite{flamm2012multiscale,diamond2013systems,leiderman2014overview,fogelson2015fluid,yesudasan2019recent,nelson2020_fibrinreview,owen2024mathematical}. Mathematical frameworks have also been developed to study fibrin polymerization on various temporal and spatial scales, where continuum and discrete, molecular dynamics approaches have been utilized to gain multiscale insight into fibrin gel formation; we refer the reader to \cite{yesudasan2019recent,nelson2020_fibrinreview} for reviews on mathematical models of fibrin polymerization. Several of these models make a continuum assumption and use ordinary (ODEs) or partial (PDEs) differential equations to study individual components of the clotting system. Studies on how the coagulation cascade is influenced by flow-mediated transport and platelet surface availability have provided mechanistic insight into various pathological states \cite{FOGELSON19981,  fogelson2012plateletcount,leiderman2011grow,miyazawa2023inhibition,tania2006inhibition}. However, these studies do not include fibrin polymerization and assume that the platelet aggregate is a continuum so clot microstructure is not studied. Despite this, computational studies \cite{leiderman2011grow} qualitatively agree with experimental studies that show development of a clot with a core-shell structure that limits transport into and out of the core \cite{stalker2013hierarchical}. To understand the clot microstructure and the impact of discrete platelet surfaces, some mathematical models of thrombus formation use a  multiscale framework. Here, platelets in the microscale are assumed to be discrete structures within a cellular Potts model \cite{xu2008multiscale} or use a Dissipating Particle Dynamics (DPD) framework to describe some volume of fluid or platelet aggregate,  where coagulation reactions in the macroscale are described by diffusion-advection-reaction PDEs \cite{tosenberger2016modelling}.

While many mathematical studies have focused on individual parts of blood clot formation, few have incorporated coagulation reactions in the fluid and on preformed platelet surfaces with fibrin polymerization and fluid transport within, into, and out of the aggregate. Some computational models have focused on understanding the biomechanics and stability of preformed, continuum clots comprised of a dense core of procoagulant platelets and fibrin mesh with a less dense shell of aggregated platelets \cite{xu2017_modelpredictions_deformation}. Others have been interested in studying the transport of solutes within continuum, preformed clots with an impermeable platelet core and a boundary layer of fixed thrombin concentration; thrombin can then transport through a permeable loose fibrin cap \cite{kim2013fibrin}. Recent studies have investigated variation in the clot microstructure to understand transport into, within, and out of the discrete platelet plug \cite{house2025imaging,Mirramezani2018_plateletpacking,tomaiuolo2014preformedclot}, showing that the densely-packed inner core traps thrombin in the intraclot gaps so solute concentrations may increase \cite{Mirramezani2018_plateletpacking} and diffusive transport decreases \cite{tomaiuolo2014preformedclot}.  Recent modeling efforts have also incorporated machine learning techniques to provide patient-specific simulations of clot microstructure formation to predict thrombus growth \cite{shankar2022three,Shankar2023_parallel_multiscale3Dthrombus,house2025imaging}.  

Here, we seek to understand how variation in platelet microstructure can impact fibrin polymerization and thus the formation of a fibrin mesh using a computational framework. Unlike in vivo or in vitro experiments,  we can gain mathematical insight into how coagulation reactions, which occur both in the fluid and on platelet surfaces, are affected by changes in the platelet microstructure. Toward this end, we develop a novel 2D computational model with preformed platelet aggregates comprised of discrete platelets with a prescribed platelet packing density and we determine flow and protein transport under physiologically relevant shear conditions. To our knowledge, this is the first mathematical modeling framework that incorporates both platelet surface reactions and coagulation reactions with a model of fibrin polymerization that can capture important aspects like gelation and branching. We also believe this is one of the first frameworks to study how surface-mediated clot reactions are impacted by the fluid volume to platelet surface area ratio within a clot. 

The computational framework we present below is fully extendable to 3D domains, but we chose a 2D approach and imagine we are simulating in vitro flow chamber experiments where there exists a preformed platelet aggregate with platelet-poor plasma flowing around the aggregate. This framework allows us to study how platelet microstructure and shear rate together modulate thrombin generation and fibrin polymerization. By varying the free fluid volume to platelet surface area ratio, we can study how spaces between the platelets could affect transport within the volume. The number of platelets per unit volume will also influence the uptake of substrates by platelets, where a large number of platelets can quickly deplete species within these smaller spaces. By quantifying the temporal and spatial metrics of clot initiation, such as thrombin concentration, fibrin development, and gelation onset, we aim to provide deeper mechanistic insights into the initial steps of thrombus formation. These insights can then be used as predictions for future experiments in which important concentrations can be quantified within a platelet aggregate. 
\begin{figure}[!h]
    \centering
 \includegraphics[width=\textwidth]{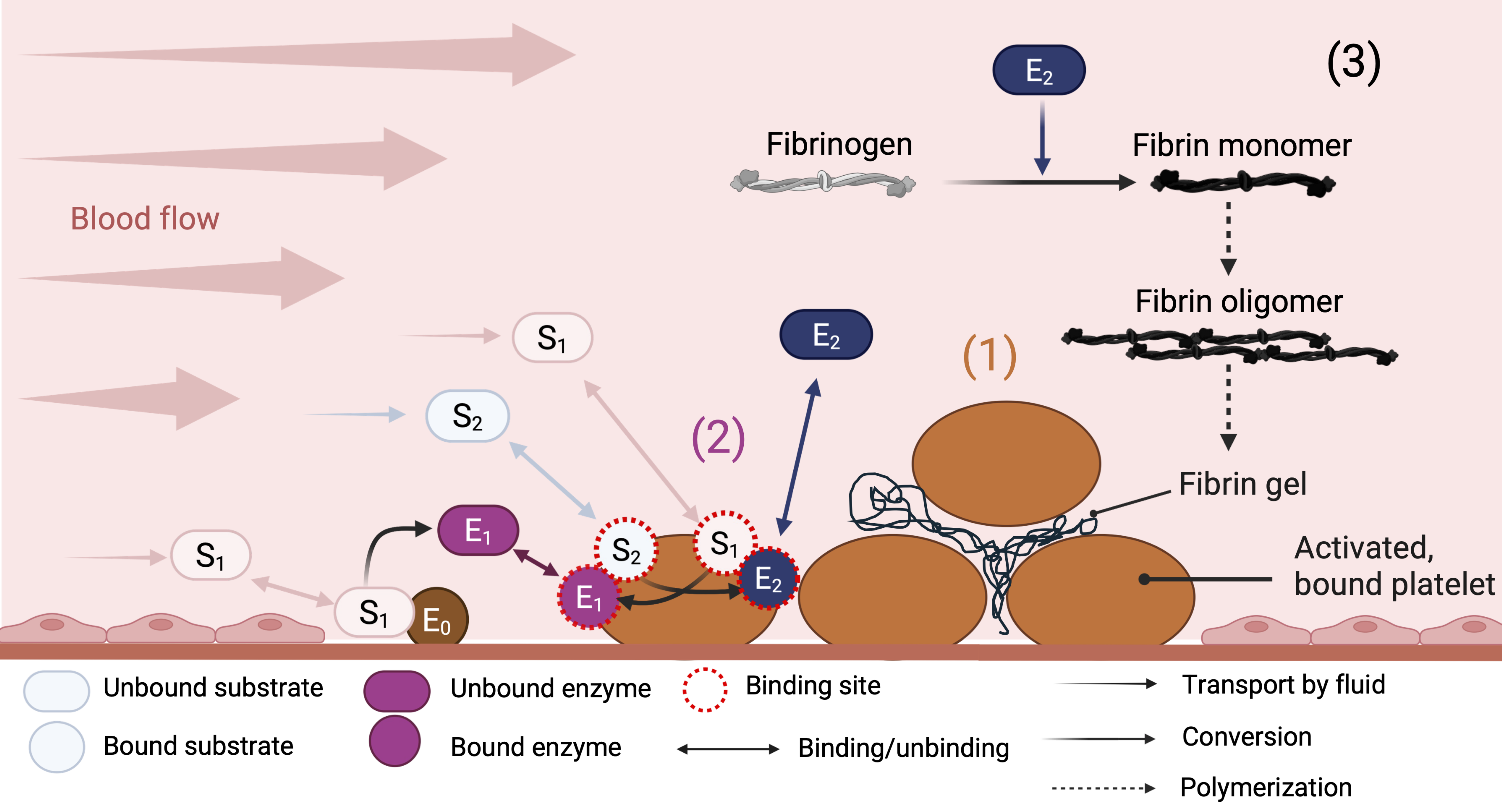}
    \caption{Schematic of mathematical framework depicting (1) the discrete platelet plug, the (2) reduced coagulation model of thrombin generation on the surfaces of platelets in the plug, and (3) the fibrin polymerization model which assumes species larger than monomer are not transported. We envision that the variables of the reduced model of coagulation roughly represent actual coagulation species $E_0$ (TF:fVIIa), $S_1$ (fX), $E_1$ (fXa), $S_2$ (prothrombin), $E_2$ (thrombin), $C_2^b=S_1^b:E_2^b$ (platelet-bound tenase fVIIIa:fIXa), and $C_1^b = S_2^b:E_1^b$ (platelet-bound prothrombinase fVa:fXa). Here,  dashed arrows represent polymerization reactions, tapered arrows represent transport, and conversion is shown with a faded arrow. Binding and unbinding of species is indicated with a double-ended arrow. Here, fluid-phase species are shown as a capsule shape,  bound species are circles, and zymogens and enzymes are represented with light and dark hued shapes, respectively. Created with BioRender.com \cite{biorender_figure}. }
    \label{fig:schematic}
\end{figure}

\section*{Materials and methods}
\label{sec:methods}

Our modeling framework consists of three main components that can affect or be affected by flow conditions: (1) a preformed platelet plug that is comprised of discrete platelets, where each activated platelet's surface contains specified numbers of binding sites for coagulation proteins; (2) a reduced model of coagulation adapted from \cite{montgomery2023clotfoam} that produces the enzyme thrombin; and (3) a model of fibrin polymerization similar to \cite{fogelson2010toward,fogelson2022development} where fibrinogen monomer is sourced upstream and is converted by thrombin to fibrin monomers that can polymerize into a fibrin gel. A schematic of these three model components under flow is shown in \Cref{fig:schematic}. This discrete-continuum modeling framework includes discrete platelets with surface concentrations of platelet-bound protein species and volumetric concentrations of other fluid-phase species. The reduced model of coagulation contains reactions among several biochemical species that are categorized as fluid-phase, platelet-bound, or subendothelium-bound. It is adapted from our more comprehensive model \cite{kuharsky2001surface}. Finally, the mathematical description of fibrin polymerization tracks over space and time the concentrations of individual fibrin monomers $c_{10}$, monomers in oligomers $\theta$, branches $B$, free reactive sites $R$, and a function $Y$ that indicates progress toward gelation \cite{fogelson2010toward,fogelson2022development}. In the subsections below and in the supplement, we describe each modeling component, and we provide parameter values used in the model in Table~\ref{tab:parameters}. 
The model uses a continuum description of fluid, coagulation factors, and fibrin species. We regard the blood as an incompressible Newtonian fluid that is governed by the Navier-Stokes equation:

\begin{equation}
\rho \left( \frac{\partial \mathbf{u}}{\partial t} + (\mathbf{u}\cdot \nabla)\mathbf{u} \right)
= -\nabla p + \mu \nabla^2 \mathbf{u},
\label{eq:momentum}
\end{equation}

\begin{equation}
    \nabla \cdot \mathbf{u} = 0.
\label{eq:conservation}
\end{equation}
where the unknowns are the fluid velocity $\textbf{u}(\textbf{x}, t)$ and fluid pressure $p(\textbf{x},t)$, and the values for the density $\rho$ and dynamic viscosity $\mu$ are given in Table \ref{tab:parameters}.  Unlike some other models that use a Darcy term in the fluid dynamics equations to model the platelet plug as a porous material \cite{leiderman2011grow,xu2017_modelpredictions_deformation,du2020clot}, here the fluid flows in the gaps between discrete stationary platelets and the presence of the platelets directly affects the fluid motion through the requirement that the fluid velocity be zero at each point of a platelet surface.  A parabolic inlet velocity profile is specified as appropriate to generate flows with wall shear rates of 0 s$^{-1}$ (no-flow), 100 s$^{-1}$, or 1000 s$^{-1}$.  A constant outlet pressure of zero is specified, and no-slip boundary conditions are applied on the portions of the domain boundary representing blood vessel walls. Flow is allowed to reach a steady-state solution before beginning to simulate the coagulation and polymerization models.

\begin{table}
\caption{Parameters and chemical species (with their corresponding variables) in the discrete platelet model, the reduced model of coagulation, and the fibrin polymerization model. Initial concentrations are given, where species without a concentration are initialized to zero.}
\centering
\footnotesize
\begin{tabular}{@{}lll@{}}
\toprule
\textbf{Parameter} & \textbf{Value}  & \textbf{Reference}\\
\midrule
Diffusion coefficient ($D$) & 5 $\times$ 10$^{-7}$ cm$^2$/s& \cite{leiderman2011grow}\\
Dynamic viscosity ($\mu$) & 1.2 mPa$\cdotp$s& \cite{baskurt_blood_2003} \\
Fluid density ($\rho$) & 1.025 g/cm$^3$ & \cite{baskurt2007handbook}\\
Platelet activation rate ($k_{\text{act}}$) & 0.05 s$^{-1}$ & Estimated\\
Platelet diameter ($P_\text{diam}$) & $2.46$ \textmu m & 
\cite{montgomery2023clotfoam}\\
Number of $P_1$ receptors per platelet ($N_1^{\text{plt}}$) & 2700 & 
\cite{montgomery2023clotfoam}\\
Number of $P_2$ receptors per platelet ($N_2^{\text{plt}}$) & 2000 & 
\cite{montgomery2023clotfoam}\\
Total $P_1$ binding site density ($N_1$) & $2.35 \times 10^{1}$ fmol/cm$^2$ & \cite{montgomery2023clotfoam}, \hyperref[S2_Text]{S2 Appendix}\\
Total $P_2$ binding site density ($N_2$) & $1.74 \times 10^{1}$ fmol/cm$^2$ & \cite{montgomery2023clotfoam}, \hyperref[S2_Text]{S2 Appendix} \\
Volumetric $C_1^b$ association rate ($k^{+,v}_{C_1}$) & $1.03 \times 10^8$ M$^{-1}$ s$^{-1}$ & \cite{montgomery2023clotfoam}\\
Volumetric $C_2^b$ association rate  ($k^{+,v}_{C_2}$) & $1.73 \times 10^7$ M$^{-1}$ s$^{-1}$ & \cite{montgomery2023clotfoam}\\
Surface $C_1^b$ association rate ($k^{+,s}_{C_1}$) & $1.08$ $\mathrm{cm^{2}\,fmol^{-1}\,s^{-1}}$& Calculated in \hyperref[S2_Text]{S2 Appendix}\\
Surface $C_2^b$ association rate ($k^{+,s}_{C_2}$) & $1.82\times 10^{-1}$ $\mathrm{cm^{2}\,fmol^{-1}\,s^{-1}}$& Calculated in \hyperref[S2_Text]{S2 Appendix}\\
\midrule
Fibrin branching rate ($k_b$) & 1.5 $\times 10^9$ M$^{-2}$s$^{-1}$ & Estimated from \cite{ryan1999structural}\\
Fibrin linking rate ($k_l$) & $8.2 \times 10^5$ M$^{-1}$s$^{-1}$ & \cite{hantgan1979assembly} \\
Fibrinogen conversion to fibrin monomer ($k_{\text{cat}}$) & 84 s$^{-1}$ & \cite{naski1991kinetic} \\
Michaelis--Menten constant ($K_m$) & 7.2 $\times 10^{-6}$ M & \cite{naski1991kinetic} \\
Gelation threshold value for $Y(\mathbf{x},t)$ ($Y_{\text{max}}$) & 5000 & Estimated\\
Thrombin inhibition by antithrombin ($k_{AT}$) & $0.0336$ s$^{-1}$ & Calculated from \cite{MIYAZAWA2023230,olson1992role} \\
\midrule
    \textbf{Chemical} & \textbf{Concentration/density} &\textbf{Reference}\\
\midrule
$E_0$ (TF:fVIIa)  &$0.02$ fmol/cm$^2$  & This study \\
$S_1$ (fX)  & 0.17 $\times 10^{-6}$ M & \cite{montgomery2023clotfoam}\\
$S_2$ (Prothrombin)  & 1.4 $\times 10^{-6}$ M & \cite{montgomery2023clotfoam}\\
$G$ (Fibrinogen)  & 5.88 $\times 10^{-6}$ M & \cite{weisel2017fibrin}\\
$E_1$ (fXa) & -- & --\\
$E_2$ (Thrombin) & --& -- \\
$c_{10}$ (Free monomers) &  --&-- \\
$B$ (Branches) & -- & --\\
$\theta$ (Monomers in oligomers) &-- & --\\
$R$ (Reactive sites) & -- &-- \\
$Y$ (Gelation indication) & -- & -- \\
\bottomrule
\label{tab:parameters}
\end{tabular}
\end{table}

\subsection*{Discrete platelet model}
\label{model:discrete}

The platelets are modeled as discrete, stationary disks with locations specified from the start of a simulation. Initially, each platelet becomes activated at a rate $k_{\text{act}}$. In the model, the effect of activation is to make binding sites on the platelet surfaces available for coagulation proteins.  These binding sites are essential for the enzymatic reactions on those surfaces. There are two classes of binding sites, $P_1$ for substrate $S_1$ and enzyme $E_1$, and $P_2$ for substrate $S_2$ and enzyme $E_2$.  The maximum number densities of the binding sites on an activated platelet are denoted $N_1^{\text{max}}$ and $N_2^{\text{max}}$, respectively (see Table \ref{tab:parameters}).  With $t$ measured from the start of a simulation, the number density of type $i$ binding sites at time $t$ is
\begin{equation}
    N_i(t)=N_{i}^{\text{max}}(1-e^{-k_{\text{act}}t}).
    \label{eq:plt_act}
\end{equation}
The binding sites are thus not available instantaneously, they increase gradually and approach their maximum over time. With $k_{\mathrm{act}}=0.05$, 95\% of binding sites become available within about 60 s. As illustrated in Figs. \ref{fig:velocity_geometry}A,B,C, we constructed three distinct pre-adhered platelet plug geometries within a 2D rectangular channel which represents a blood vessel. The geometries differed in the uniform interplatelet gap size used; 0.1 \textmu m (dense), 0.5 \textmu m (medium), and 1 \textmu m (loose). The platelets have diameter 2.46 \textmu m and a consistent spacing of 0.5 \textmu m is maintained between the lowest row of platelets and the bottom vessel wall. The platelet plug geometries are determined to achieve consistent overall dimensions, forming a half ellipse shape with maximum height of 17 \textmu m and width of 54 \textmu m. \Cref{tab:plt_char} contains the characteristics of each geometry.  Each platelet arrangement was generated using LAMMPS molecular dynamics software, and was subsequently meshed in Ansys Mechanical, where a general body element size of 2 \textmu m is specified for use far from the platelets and refined near the platelets to conform with a mesh element size of 0.1 \textmu m for platelet surfaces and 1 \textmu m for the lower wall surface.

\subsection*{Reduced coagulation model}
\label{model:coag}

The coagulation reaction system is complex with multiple enzyme activation steps, feedback reactions, and inhibitors.  Previous models, including our own \cite{  diamond2013systems,leiderman2014overview,tania2006inhibition,leiderman2011grow,miyazawa2023inhibition}, have incorporated up to approximately 140 species.  Here, we use a reduced model which we believe adequately captures the salient features of thrombin production for our current purposes. Importantly, the reduced model involves biochemical species that can be subendothelium-bound (SE-bound), platelet-bound, or fluid-phase. The critical reactions happen on the subendothelial surface or platelet surfaces, and products of reactions on one surface must move through the fluid to other surfaces to participate in other reactions. So, diffusive and advective transport play important roles in providing substrates for the surface-phase enzyme reactions, allowing enzymes produced on one surface to move to a second surface, and in carrying fluid-phase enzymes out of the platelet aggregate and then downstream. The model we use is adapted from that of Montgomery et al. \cite{montgomery2023clotfoam} and is similar to the earlier model of Fogelson and Kuharsky \cite{FOGELSON19981}.

A schematic of the reduced model reactions is given in \Cref{fig:schematic}.  The model involves enzyme $E_0$ on the SE that roughly corresponds to the TF:fVIIa complex.  It forms a complex with fluid-phase substrate $S_1$ and converts the substrate into fluid-phase enzyme $E_1$, which roughly represents fXa. Both $S_1$ and $E_1$ can bind to $P_1$ receptors on the   surfaces of activated platelets to form the platelet-bound species   $S_1^b$ and $E_1^b$, respectively.  Similarly, fluid-phase substrate   $S_2$ and enzyme $E_2$ can bind to $P_2$ receptors on activated platelets to form the platelet-bound species $S_2^b$ and $E_2^b$. $S_2$ corresponds roughly to prothrombin and $E_1^b$ is analogous to prothrombinase in that $S_2^b$ and $E_1^b$ can bind to form complex $C_1^b$ that then activates the $S_2^b$ in that complex to $E_2^b$. $E_2^b$ can then bind with $S_1^b$ to form the $C_2^b$ complex which converts its $S_1^b$ into enzyme $E_1^b$. The feedback activation of platelet-bound $S_1^b$ by $E_2^b$ is a simplified version of the feedback activity of real thrombin in increasing the concentration of the platelet-bound tenase complex which produces fXa on activated platelet surfaces.  Both enzymes $E_1^b$ and $E_2^b$ can unbind from platelets becoming fluid-phase $E_1$ and $E_2$, respectively, and can then bind to other platelets. Fluid-phase $E_2$ is inhibited by antithrombin  (AT) through a first-order reaction.  Importantly, it can also react with fibrinogen in the fluid to produce fibrin monomers, as described below. 

The concentrations of the fluid-phase species $S_1$, $E_1$, $S_2$, and $E_2$ evolve according to partial differential equations that describe their transport by advection and diffusion.  Boundary conditions for these species couple their motion with their binding to and unbinding from surfaces.  The surface densities of the platelet-bound species $S_1^b$, $E_1^b$, $S_2^b$, and $E_2^b$ evolve according to ordinary differential equations at each point on each activated platelet's surface.  These ordinary differential equations reflect the effect of the reactions that influence the rate of change of the surface densities. We ignore the possibility of diffusive transport on the platelet surfaces.

\begin{table}[htbp]
\centering
\caption{Platelet plug characteristics for different packing densities in predefined region.}
\begin{tabular}{@{}lcccc@{}}
\toprule
\textbf{Characteristic} & \textbf{Dense} & \textbf{Medium} & \textbf{Loose} \\
\midrule
Number of platelets [--] & 114   & 79    & 63    \\
Concentration [platelets/mm$^3$] & $6.84 \times 10^{7}$ & $4.44 \times 10^{7}$ & $2.78 \times 10^{7}$ \\
Area occupied (2D) [\%] & 83.89 & 62.86 & 46.02 \\
Free-space (2D) [\%]    & 16.12 & 37.14 & 53.98 \\
Free-space (2D) [\textmu m$^{2}$] &247.12 &417.98  & 492.81\\
Gap size [\textmu m] & 0.1& 0.5 & 1.0\\
Average intraplug velocity [\textmu m/s] & 28.12 & 76.55& 110.76 \\
Average intraplug Peclet number [--] &0.056 & 0.77 & 2.22 \\
\bottomrule
\end{tabular}
\label{tab:plt_char}
\end{table}

The complete set of differential equations of the reduced coagulation model is given in \nameref{S1_text}.  Here, we use two of those equations and the corresponding boundary conditions to illustrate the form of this part of the model.  The concentration of the fluid-phase substrate $S_1$ satisfies the advection-diffusion equation
\begin{equation}
\frac{\partial S_1}{\partial t} = -\nabla \cdot \left( \mathbf{u} S_1 - D \nabla S_1 \right).
\label{eq:S1pde}
\end{equation}
Because $S_1$ can bind to the SE-bound enzyme $E_0$ and to platelet-surface receptors, $P_1$, $S_1$ satisfies boundary conditions both on the SE and on the surfaces of platelets.  The boundary conditions are obtained by equating the diffusive flux of $S_1$ to a point on the boundary with the net rate of binding of $S_1$ to receptors on that surface:
\begin{equation}
- D \, \hat{n} \cdot \nabla S_{1} \Big|_{\partial \Omega_{\text{inj}}} = -k_{C_0}^{+} S_1 E_0^b + k_{C_0}^- C_0^b,
\label{eq:S1BCbottom}
\end{equation}
at all points of the subendothelium and 
\begin{equation}
- D \, \hat{n} \cdot \nabla S_{1} \Big|_{\partial \Omega_{\text{pla}}}
= -k_{S_{1}}^{+} \left\{ N_{1} - \left( S_{1}^{b} + E_{1}^{b} + C_{1}^b + C_{2}^b \right) \right\} S_{1} 
+ k_{S_{1}}^{-} S_{1}^{b}
\label{eq:S1BCplatelet}
\end{equation}
at all points of each activated platelet surface.  In this equation, the quantity
$N_{1} - \left( S_{1}^{b} + E_{1}^{b} + C_{1} + C_{2} \right)$ is the surface density of \textit{unoccupied} $P_1$ binding sites for $S_1$ and $E_1$ on the platelet's surface.  The concentration $S_1$ satisfies zero diffusive-flux boundary conditions at points of the uninjured vessel walls, is set to its bulk plasma concentration at points on the inlet to the computational domain, and satisfies a zero diffusive-flux boundary condition at points of the outlet of the domain.

The surface density of the platelet-bound substrate $S_1^b$ satisfies an ordinary differential equation at all points of each activated platelet surface 
\begin{equation}
\frac{\partial S_1^b}{\partial t} = k_{S_1}^{+} \left\{ N_1 - \left( S_1^b + E_1^b + C_1^b + C_2^b \right) \right\} S_1 - k_{S_1}^- S_1^b  -  k_{C_2}^{+,s} S_1^b E_2^b + k_{C_2}^- C_2^b.  
\label{eq:S1bode}
\end{equation}
Previous models including reactions between pairs of platelet-bound species, have treated these as volumetric reactions with the platelet-bound species quantified by their effective volume concentrations \cite{kuharsky2001surface,leiderman2011grow}. Similarly, experiments measuring the rates of protein interactions of platelet-bound species report ``apparent" reaction rates, treating the reactants as distributed in the solution volume \cite{colman_walsh_chapter,Walsh2004-kf,colman_mann_chapter}.  In the current model, platelet-bound species are found on the surfaces of discrete platelets and are quantified as surface densities (amount per area).  This necessitates our converting the apparent volumetric rate constants, e.g., $k_{C_2}^{+,v}$ into appropriate equivalents, $k_{C_2}^{+,s}$, for surface reactions between pairs of surface-bound species.  To do this we require that the total volumetric rate of reaction for a representative volume matches the total rate of surface-bound reactions for all the platelet surfaces in that volume. See \hyperref[S2_Text]{S2 Appendix} for details.

\subsection*{Fibrin polymerization model} 
\label{sec:fibrinmodel}

The model of fibrin polymerization used in this paper is inspired by that in \cite{fogelson2010toward} and describes reactions in which fibrin monomers can polymerize to form branched fibrin oligomers. We view a fibrin monomer as a linear molecule with two half-monomer domains, and each half-monomer domain is a free reaction site; that is, each fibrin monomer has two free reaction sites. Each free site can participate in two types of reactions: linear polymerization in which two free sites come together to form a linear
link, and branch formation in which three free sites come together to form a trimolecular “branch”.  Branch formation produces species with more than two reaction sites and sufficient branch formation is required for gelation to occur \cite{fogelson2022development}. Each fibrin oligomer, formed by these two reactions, can be described by two indices, $m$ and $b$, where $b$ is the number of branches and $m + 2b$ is the total number of monomers in the oligomer. Therefore, a free (unbound) fibrin monomer is denoted $C_{10}$ and an oligomer with $m + 2b$ monomers and $b$ branches is denoted as $C_{mb}$. With this notation, the formation of a linear link can be described as:
\begin{equation}
    C_{m_1,b_1} + C_{m_2,b_2} \xrightarrow[]{k_l}C_{m_1+m_2, b_1+b_2},
    \label{eq:link}
\end{equation}
and the formation of a branch can be written as 
\begin{equation}
    C_{m_1,b_1} + C_{m_2,b_2} +  C_{m_3,b_3}  \xrightarrow[]{k_b}C_{m_1+m_2 + m_3 - 2, b_1+b_2+b_3+1}.
    \label{eq:branch}
\end{equation}
The concentration of oligomers of type $C_{mb}$ is denoted $c_{mb}(\textbf{x}, t)$ and its time rate of change in Equation (1) in \nameref{S3_text}.  Those equations account for changes in concentrations due to polymerization reactions, due to there being a source of monomer at locations at which both fibrinogen and thrombin are present, and due to transport. For simplicity, we assume that only fibrin monomers move by advection with the fluid and by diffusion; larger oligomers are regarded as being stationary. More details and justification for this assumption can be found in the \nameref{sec:discussion}. The model also tracks the concentration of fibrinogen, denoted $G(\textbf{x},t)$ and its conversion to fibrin monomers by the enzyme thrombin ($E_2(\textbf{x},t)$) according to Michaelis--Menten kinetics.  To the extent possible, we use parameter values taken from the experimental literature (see Table \ref{tab:parameters}). There is no value in the literature for the branching rate parameter $k_b$, and we have used experimental data \cite{ryan1999structural} to estimate this parameter in \nameref{S3_text} (see Fig S1).  As in \cite{fogelson2010toward}, we exclude self-interactions between reaction sites on an oligomer so no cycles may form.  Consequently, the number of free reaction sites on a $C_{mb}$ oligomer is $b+2$.  We assume that the rates of reactions among oligomers as included in Equation (1) are proportional to the number of free reaction sites on each of the reacting species.

We are interested in the spatial-temporal dynamics of several quantities that can be defined in terms of the set of oligomer concentrations.  These include the total concentration of free reaction sites $R(\textbf{x},t)$, the total concentrations of monomers in all oligomers $\theta(\textbf{x},t)$, the total concentration of branches $B(\textbf{x},t)$, and the average number of monomers in an oligomer $A(\textbf{x},t)$ which we also refer to as the average cluster size.  The definitions of these quantities are
\begin{equation}
    R(\mathbf{x},t) = \sum_{m,b} (b+2)c_{mb}(\mathbf{x},t),
\end{equation}
\begin{equation}
    \theta(\mathbf{x},t) = \sum_{m,b}(m+2b)c_{mb}(\mathbf{x},t),
\end{equation}
\begin{equation}
    B(\mathbf{x},t)  = \sum_{m,b} b c_{mb}(\mathbf{x},t),
\end{equation}
and
\begin{equation}
\label{eq:average}
    A(\mathbf{x},t)= \frac{1}{\theta}\sum_{m,b} (m+2b)^2c_{mb}(\mathbf{x},t).
\end{equation}
Following the arguments in \cite{fogelson2010toward,fogelson2022development,nelson2023towards}, we define the onset of gelation at location $\textbf{x}$ as the time, $t_{gel}(\textbf{x})$ at which the $A(\textbf{x},t) \rightarrow \infty$.  Loosely speaking, gelation occurs when the average cluster size becomes infinite and since oligomers with more than two free sites are required to form a gel \cite{ziff1980kinetics}, sufficient branching is necessary.  

As defined, we need to know all the concentrations $c_{mb}$ to calculate the quantities of interest.  The differential equations for these concentrations form a doubly-infinite set precluding their solution. Instead, using a moment-generating function approach \cite{fogelson2022development,nelson2023towards}, we derive from Equation (1) in \nameref{S3_text} a small set of differential equations for select zeroth, first, and second moments of the oligomer distribution, and we use the fact that the quantities of interest defined above can be expressed as linear combinations of these moments, to derive a small closed set of differential equations directly for those quantities.  Rather than tracking $A(\textbf{x},t)$, it is convenient to track $Y(\textbf{x},t) = M_{0 2}(\textbf{x},t) - R(\textbf{x},t)$ which becomes infinite at a specific time if and only if $A(\textbf{x},t)$ becomes infinite at that time \cite{nelson2023towards}.  The quantity $M_{02} = \sum_{m,b} (b+2)(b+1) c_{mb}$ is a second moment of the oligomer distributions with respect to the number of branches $b$. The differential equations for $R$, $\theta$, $B$, and $Y$ are
\begin{eqnarray}
    \frac{\partial R(\mathbf{x},t)}{\partial t} &=& -k_lR(\mathbf{x},t)^2 - \frac{k_b}{2}R(\mathbf{x},t)^3 \label{eq:r} \\
    &+& 2\left(-\mathbf{u} \cdot \nabla c_{10}(\mathbf{x},t) + D\Delta c_{10}(\mathbf{x},t) + k_{\text{cat}}E_2(\mathbf{x},t)\frac{G(\mathbf{x},t)}{K_m + G(\mathbf{x},t)}\right),\nonumber\\
   \frac{\partial B(\mathbf{x},t)}{\partial t} & = & \frac{k_b}{6}R(\mathbf{x},t)^3,
    \label{eq:b} \\
 \frac{\partial \theta(\mathbf{x},t)}{\partial t} &=& -\mathbf{u}\cdot\nabla c_{10}(\mathbf{x},t) + D\Delta c_{10}(\mathbf{x},t) + k_{\text{cat}} E_2(\mathbf{x},t)\frac{G(\mathbf{x},t)}{K_m + G(\mathbf{x},t)},
 \label{eq:thetas} \\
     \frac{\partial Y(\mathbf{x},t)}{\partial t} &=& k_l Y(\mathbf{x},t)^2 + k_b R(\mathbf{x},t)\left(\frac{R(\mathbf{x},t)^2}{2} + R(\mathbf{x},t)Y(\mathbf{x},t) + Y(\mathbf{x},t)^2\right).
     \label{eq:y}
\end{eqnarray}  
These equations and differential equations for the fibrinogen concentration $G$ and free fibrin monomer concentration $c_{10}$ constitute the fibrin polymerization part of our model.  The former is
\begin{equation}
    \frac{\partial G(\mathbf{x},t)}{\partial t} = -\mathbf{u}\cdot\nabla G(\mathbf{x},t) + D\Delta G(\mathbf{x},t) - k_{\text{cat}} E_2(\mathbf{x},t)\frac{G(\mathbf{x},t)}{K_m + G(\mathbf{x},t)},
    \label{eq:g}
\end{equation}
and the latter is given in Equation\ \eqref{eq:c10} below.  Note that the transport terms in Equations\ \eqref{eq:r} and \eqref{eq:thetas} describe monomer transport.  By simple changes of variables, we can remove monomer transport terms from all but Equation\ \eqref{eq:c10} (see \nameref{S3_text} for details).

It is important that Equations\ \eqref{eq:r}-\eqref{eq:y} are valid at location $\textbf{x}$ only until the time at which a gel forms there.  We therefore need to detect when this occurs, and to specify how concentrations evolve post gelation.  Towards this end, we introduce a gelation indication function $I(\textbf{x},t)$ defined to equal 1 if a gel has formed at location $\textbf{x}$ by time $t$, and to equal 0 otherwise.  In preliminary computational experiments, we found that we could define a reasonable gelation threshold, namely gelation occurs at $\textbf{x}$ at the first time $t_{gel}(\textbf{x})$ at which $Y(\textbf{x},t)$ exceeds $Y_{\text{max}} =  $ 5000 nM.  Hence, $I(\textbf{x},t)$ is defined by
\begin{equation}
    I(\mathbf{x},t) = \begin{cases}
        0 & \text{ if } t<t_{gel}(\mathbf{x})\\
        1 & \text{ if } t\geq t_{gel}(\mathbf{x})\\ 
    \end{cases}.
\end{equation}
Once gelation has occurred at $\textbf{x}$, we assume that all polymerization reactions cease there, and that the only reactions that can occur there are diffusive and advective transport of fibrin monomers and the conversion of fibrinogen to fibrin monomers by thrombin. We also assume that the fluid dynamics equations are unaffected by the formation of fibrin polymer, so transport of monomeric species is not impacted by the appearance of a gel.  With these assumptions, the equation for $c_{10}(\textbf{x},t)$ can be written
\begin{equation}
\begin{split}
    \frac{\partial c_{10}(\mathbf{x},t)}{\partial t} &= -\mathbf{u} \cdot \nabla c_{10}(\mathbf{x},t) + D\Delta c_{10}(\mathbf{x},t) + k_{\text{cat}}E_2(\mathbf{x},t)\frac{G(\mathbf{x},t)}{K_m + G(\mathbf{x},t)}\\
    &-\delta_{I(\mathbf{x},t),0}\big(2k_l R(\mathbf{x},t) + k_b R^2(\mathbf{x},t)\big)c_{10}(\mathbf{x},t),
    \end{split}
     \label{eq:c10}
\end{equation}
where $\delta_{i,j}$ is the Kronecker delta.   

The fibrinogen concentration at the inlet to the domain is set to its plasma value and  is assumed to satisfy zero diffusive flux boundary conditions at all points of the vessel wall and the outlet.  The concentrations of all fibrin species are assumed to satisfy zero diffusive flux boundary conditions at all points of the domain boundary.

\subsection*{Simulations}

To simulate the dynamics of Equations \eqref{eq:momentum} and \eqref{eq:conservation}, we utilized ANSYS Fluent 2023R2.  Pressure–velocity coupling employed the coupled algorithm option; pressure was discretized with a second-order scheme, and time integration used the second-order implicit formulation option.  For fluid-phase biochemical species ($S_1$, $E_1$, $S_2$, $E_2$, $G$, $c_{10}$) that undergo transport, the corresponding reaction diffusion advection equation (e.g. Equation\ \eqref{eq:S1pde}) along with the appropriate boundary conditions (Equations\ \eqref{eq:S1BCbottom}--\eqref{eq:S1BCplatelet}) was solved in ANSYS Fluent using the user-defined scalar framework. For transported species, Fluent assembles the convection and diffusion operators using the local velocity field together with a constant diffusivity $D$, with all spatial terms discretized using a second-order upwind scheme and gradients evaluated with the least-squares cell-based method. Fluid-phase species that are not transported ($R$, $B$, $\theta$, $Y$) were assigned zero diffusivity and convection was disabled, such that their evolution reduces to spatially local reaction ODEs, which Fluent solves implicitly.  The concentrations of surface-bound species satisfy ODEs (e.g., Equation\ \eqref{eq:S1bode}) at each point of the appropriate surface, and were treated through a boundary ODE system, implemented with a user-defined function. At each time step, the procedure retrieves the current boundary values, evaluates the reaction terms, and performs an explicit forward-Euler update using the solver’s time-step size. This ensures consistent coupling between surface reactions and bulk coagulation dynamics throughout the transient simulation.  Discussion of the mesh and time-step independence of the numerical solutions as well as spatial and temporal convergence studies appear in \nameref{S4_text}. The simulations are performed for a physical time of one or two minutes to capture early blood clot formation. This choice of two minutes is chosen as a time period where platelet-driven, fibrin-mediated clot contraction is not expected to be significant \cite{tutwiler2016kinetics}. 

  \begin{figure}[!h]
        \centering
        \includegraphics[width=\textwidth]{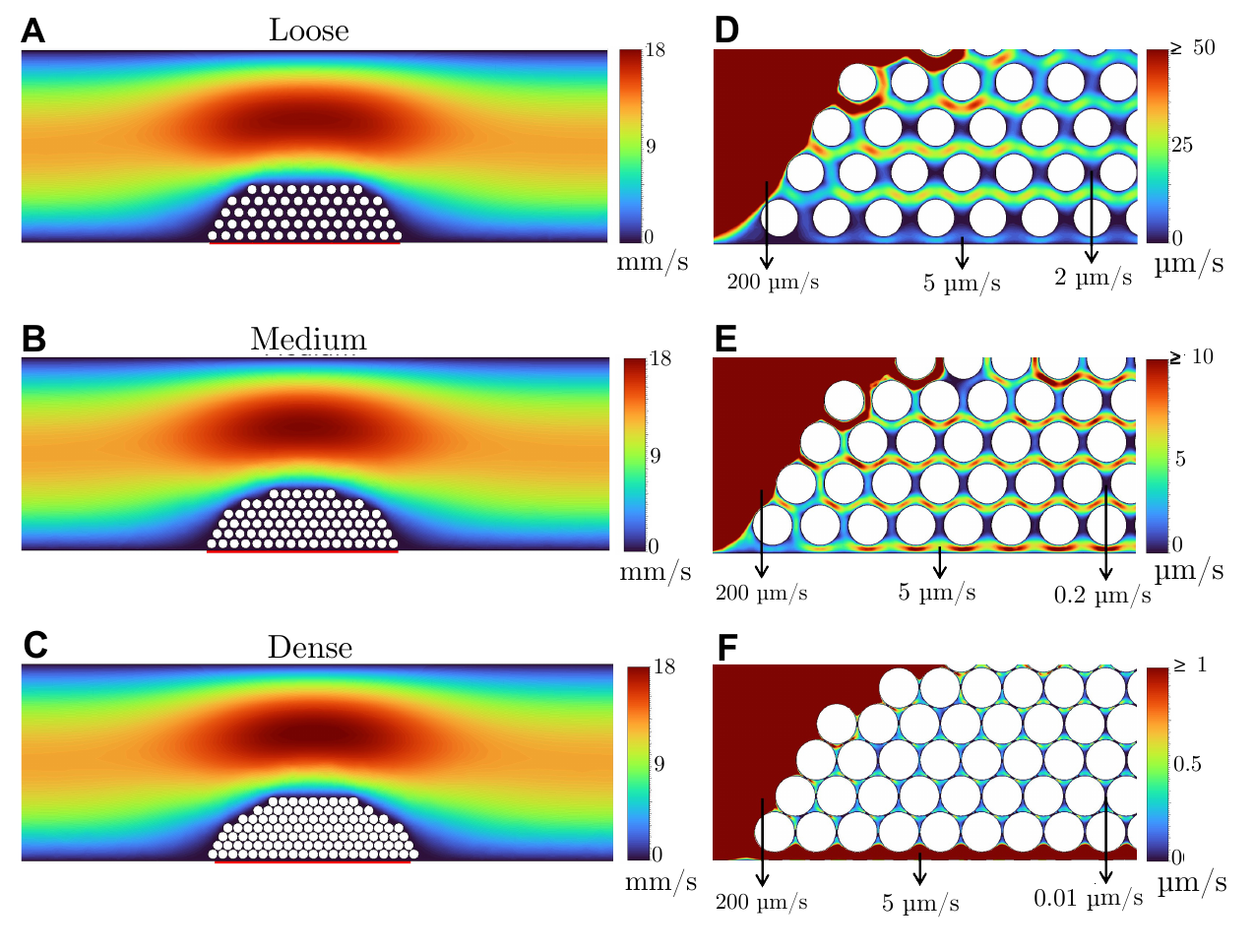}    
        \caption{Velocity profiles for each platelet plug for an initial wall shear rate of 1000 s$^{-1}$. Whole domain velocity profiles for (A) loose, (B) medium, and (C) dense platelet plug geometry. Velocity profiles for intraplug velocities for (D) loose, (E) medium, and (F) dense platelet plug geometries. Colors represent computed velocity values and note the difference in color bar ranges/units for each subfigure. The red strip of wall below the platelet aggregate represents a 50~\textmu m length segment of the vessel wall where $E_0$ is nonzero. Note the differences in velocity ranges for (D), (E), and (F). }
        \label{fig:velocity_geometry}
        \end{figure}

\begin{table}
\caption{Reactions on subendothelium and platelet surface (kinetic constants in solution and on the surface), with relevant parameter and parameter values given. Surface reaction rates are calculated from volumetric reaction rates in \nameref{S2_Text}. All reaction rates except $k_{C_i}^{+,s}$ are taken from \cite{montgomery2023clotfoam}.}
\centering
\footnotesize
\begin{tabular}{lcccccc}
\toprule
Activation (of --, by --) & Complex & Product  & $k^{+,s}_{C_i}$($\mathrm{cm^{2}\,fmol^{-1}\,s^{-1}}$) &
$k^{+,v}_{C_i}$ (M\textsuperscript{$-1$} s\textsuperscript{$-1$}) & 
$k^{-}_{C_i}$ (s\textsuperscript{$-1$}) &
$k^{\text{cat}}_{C_i}$ (s\textsuperscript{$-1$}) \\
\midrule
$(S_{1},E_{0}^b)$               & $C_{0}^b$ & $E_{1}$       &  --&  {$8.95\times10^{6}$}& $1.0$ & $1.15$ \\
$(S_{2}^{b},E_{1}^{b})$       & $C_{1}^b$ & $E_{2}^{b}$  & $1.08$& $1.03 \times 10^8$& $1.0$ & $30.0$ \\
$(S_{1}^{b},E_{2}^{b})$       & $C_{2}^b$ & $E_{1}^{b}$  &  $1.82\times10^{-1}$& $1.73 \times 10^8$ & $1.0$ & $0.23$ \\
\midrule
Binding of & Reactants & Product &
\multicolumn{2}{c}{$k^{+}_{*_i}$ (M\textsuperscript{$-1$} s\textsuperscript{$-1$})} &
\multicolumn{2}{c}{$k^{-}_{*_i}$ (s\textsuperscript{$-1$}) }  \\
\midrule 
$S_{1}$ (X)  & $S_{1},P_{1}$ & $S_{1}^{b}$  &\multicolumn{2}{c}{$5.7\times10^{7}$} & \multicolumn{2}{c}{$0.17$ }           \\
$E_{1}$ (Xa) & $E_{1},P_{1}$ & $E_{1}^{b}$ &\multicolumn{2}{c}{$1.0\times10^{7}$} & \multicolumn{2}{c}{$2.5\times10^{-2}$} \\
$S_{2}$ (II) & $S_{2},P_{2}$ & $S_{2}^{b}$ & \multicolumn{2}{c}{$1.0\times10^{7}$} &   \multicolumn{2}{c}{$5.9$}             \\
$E_{2}$ (IIa)& $E_{2},P_{2}$ & $E_{2}^{b}$&\multicolumn{2}{c}{$1.0\times10^{7}$} & \multicolumn{2}{c}{$5.9$}             \\
\bottomrule
\end{tabular}
\label{tab:binding}
\end{table}

     \section*{Results}

In this section, we show how varying the platelet packing density impacts the dynamics of both the coagulation and fibrin polymerization models.  We look at three platelet plug configurations, defined as dense, medium, and loose; details about the platelet plug geometry are given in Table~\ref{tab:plt_char}. Since the intraplug microstructure impacts transport within and around the region of the platelet plug, we also investigate how the formation of a fibrin gel is impacted by different shear conditions. All parameter values and initial conditions used for simulating the reduced coagulation model and the fibrin polymerization model are shown in Tables~\ref{tab:parameters} and \ref{tab:binding}.

\subsection*{Increased platelet packing density decreases fluid velocity within platelet plug}
\Cref{fig:velocity_geometry} shows how the fluid velocity is impacted by the presence of a platelet plug and how the velocity within the platelet microstructure is modified for different packing densities. For all three configurations in Figures~\ref{fig:velocity_geometry}A,B,C, the maximum velocity magnitude in the center of the vessel is 18 mm/s and just upstream from the platelet aggregate and about 1 platelet diameter from the vessel wall, velocity magnitudes are around 200 \textmu m/s.  Figures~\ref{fig:velocity_geometry}D,E,F show how the platelet packing density influences the velocity inside the platelet plug interior. The velocities within the plugs range between 2 and 50 \textmu m/s for the loose plug, and between 0.01 and 0.1 \textmu m/s for the dense one.

\subsection*{Platelet packing density impacts volume-integrated amounts of enzymes and fibrin}

\begin{figure}[!h]
        \centering    
        \includegraphics[width=
        \textwidth]{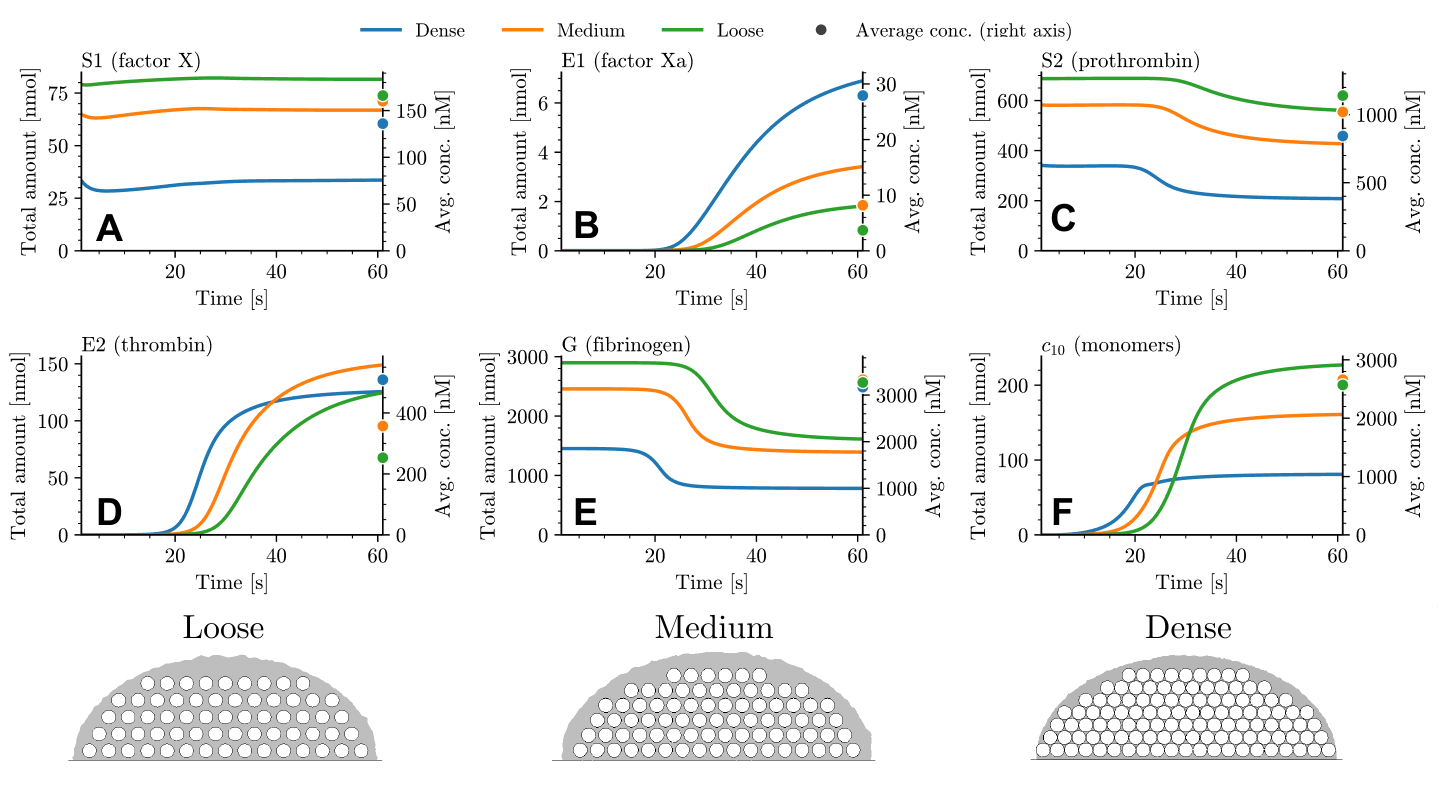}
        \caption{Effect of platelet-plug packing density on volume-integrated species concentrations, at a shear rate of 1000 s$^{-1}$. Top: Time courses (0–61 s) of the total amounts of (A) $S_1$ (factor X), (B) $E_1$ (factor Xa), (C) $S_2$ (prothrombin), (D) $E_2$ (thrombin), (E) $G$ fibrinogen, and (F) $c_{10}$ fibrin monomers for the three geometries, denoted with different colors. Each marker on the right $y$-axis corresponds to the average concentration at 60 s for a given geometry. Bottom: Schematic half-ellipsoidal plugs with identical outer dimensions but different packing (left to right: loose, medium, dense). The grey region represents the intraplug fluid domain; all totals are computed by integrating concentrations over this region only. See \Cref{tab:plt_char} for geometric characteristics.}
        \label{fig:total_conc}
        \end{figure}

Figure\ \ref{fig:total_conc} gives a spatially-integrated picture of how variations in the platelet packing density influence the consumption and production of the coagulation species and the production of fibrin monomers for shear rate 1000 s$^{-1}$.  For each fluid-phase species and time, we integrate the species' concentration over the fluid portion of the plug, indicated in gray in the bottom row of the figure, to determine the total amount of that species within the plug as a function of time, as plotted on the left $y$ axis. On the right $y$ axis, we show with a marker the average species concentration for that platelet configuration at 60 s. Figure\ \ref{fig:total_conc}A,C,E show the amounts of substrates $S_1$ and $S_2$ representing factor X and prothrombin, respectively, and of $G$ representing fibrinogen for the three plug geometries.  We see that the amount of each substrate decreases as the packing density increases; this is a direct consequence of the reduction of interplatelet space (54\%, 37\%, and 16\% of the total plug area for the loose, medium, and dense plugs, see \Cref{tab:plt_char}). However, the total amount of each of the substrates ($S_1,\,S_2, \,G$) approaches a quasi-steady value consistent with their consumption being balanced by replenishment due to transport.   We see in Fig.\ \ref{fig:total_conc}B,D,F that after an initial latent period of 15--30 s, the amounts of the products $E_1$ (factor Xa), $E_2$ (thrombin), and $c_{10}$ (fibrin monomer) begin to grow, with the onset of the accumulation occurring sooner for denser plug geometries. For $E_1$, at each time the total amount is highest for the dense plug and lowest for the loose one, despite the fact that the amount of substrate $S_1$ runs in the opposite order.  For $E_2$, the amount rises first and fastest for the dense plug, but the rise then slows, and by 60 s, the amounts for the medium and loose plugs have matched or exceeded that for the dense one.  For $c_{10}$, the denser the plug, the sooner the rapid rise begins and the sooner it slows, so that by 60 s, there is substantially more fibrin monomer for the loose plug than for the medium or dense ones. Finally, the average concentration of the substrates and fibrin monomer is approximately the same for each of the platelet configurations at time 60 s, whereas the concentration of products generated on the platelet surface differ for varying platelet packing densities. In particular, the concentration of $E_1$ in the dense configuration is more than an order of magnitude larger than for the loose configuration.

\subsection*{Location of fibrin gel initiation depends on platelet packing density and shear rate}
Thrombin produced by the reduced coagulation model converts fibrinogen into fibrin monomers, which polymerize and eventually form an insoluble gel. As described in the \nameref{sec:methods} section, we use a gelation indicator variable, $I(\mathbf{x},t)$, to track the spatial and temporal dynamics of the gel.  A value $I(\mathbf{x},t)=1$ indicates that fibrin has formed a gel at location $\textbf{x}$ prior to time $t$.

   \begin{figure}[h!] 
   \centering
      \includegraphics[width=\textwidth]{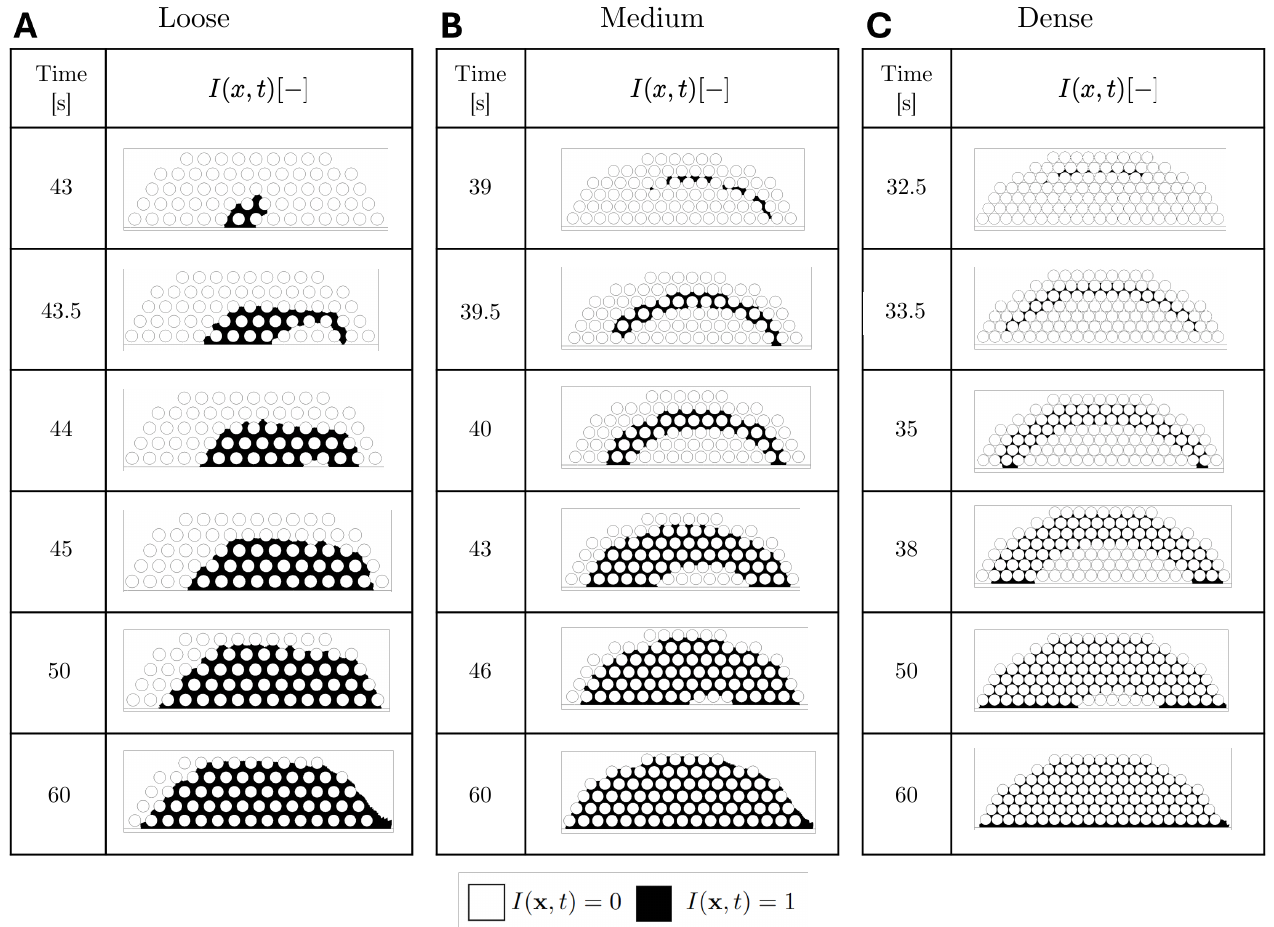}
        \caption{Comparison of gelation indicator $I(\mathbf{x},t)$  between (A) loose, (B) medium, and (C) dense platelet plugs, at shear rate = 1000 s$^{-1}$. Black regions in the plot correspond to spatial locations where the gelation indicator variable  $I(\mathbf{x},t) = 1$. Note the difference in time across different platelet configurations.  }
        \label{fig:gelation_config}
        \end{figure}

\Cref{fig:gelation_config} shows how the location of the  start and subsequent development of fibrin gelation (black, corresponding to indication variable $I(\mathbf{x},t)= 1$) is impacted by the platelet-packing configuration for shear rate 1000 s$^{-1}$ up to 60 seconds. Fibrin gel formation starts earliest for the dense plug in \Cref{fig:gelation_config}C, followed by the medium plug in \Cref{fig:gelation_config}B, and then the loose plug in \Cref{fig:gelation_config}A. Gelation starts  at the bottom of the plug near the subendothelium in the loose configuration, but starts at the periphery for the medium and dense plugs. Over time, the gel front moves toward the periphery  of the plug for the loose configuration. In contrast, for the medium and dense plugs, the gel front moves towards the plug's interior. In each case, once gelation begins, it spreads quickly until (almost) the whole platelet plug is covered. Finally, we see one effect of flow in that the gel location is skewed slightly downstream, with the greatest skew seen for the loose plug.

\subsection*{Higher shear rate decreases fibrin area accumulation}

\begin{figure}[!h]
        \centering     \includegraphics[width=\textwidth]{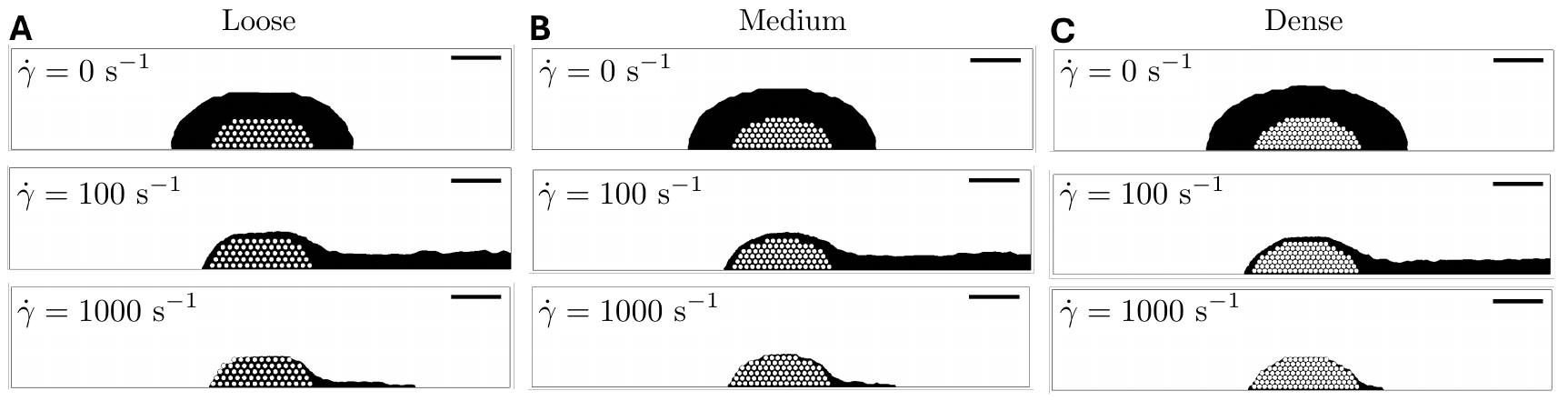}
        \caption{Fibrin gel area after 120 s of simulation time for (A) the loose, (B) medium, and (C) dense platelet plugs. Each subfigure corresponds to a different shear condition where (top) no shear, (middle) medium shear, $\dot \gamma = 100$ s$^{-1}$, and (bottom) high shear, $\dot \gamma = 1000$ s$^{-1}$ are shown. Horizontal bar corresponds to 25 \textmu m.}
        \label{fig:geometries}
        \end{figure}

To see the effect of longer time, \Cref{fig:geometries} shows the fibrin gel area at 120 seconds.  We find that without flow each platelet packing density yields a fibrin gel that extends symmetrically beyond the platelet plug in all directions. With nonzero shear rate, we see that the fibrin gel area after two minutes is restricted almost exclusively to downstream of the platelet plug, with more area for shear rate $100$ s$^{-1}$ than for shear rate 1000 s$^{-1}$ for all platelet plugs.  For a given flow, the gel area downstream from the platelet aggregate decreases as the packing density increases.  Some fibrin gel is seen upstream and above the platelet region for shear rate $100$ s$^{-1}$, but there is almost no fibrin gel in these locations for shear rate 1000 s$^{-1}$. In \nameref{sec:2D_trans_fibrin},  we show how the fibrin gel area changes over time from 30 seconds to 120 seconds in Fig S5. We find that the flow conditions and the platelet plug microstructure impact the time at which a gel first appears, as well as which plug yields the largest fibrin gel area over time. In particular, as shear rate increases, the total fibrin gel area decreases for all plugs. Without flow, the dense plug has the largest gel area by 120 seconds, followed by the medium and loose plugs. However, for both shear rate $\dot\gamma = 100$ s$^{-1}$ and $\dot\gamma =$ 1000 s$^{-1}$, we find that fibrin gel area at 120 seconds instead decreases as platelet packing density increases, with the loose plug resulting in the largest fibrin gel area.

\subsection*{Increased platelet packing density results in a decrease in replenishment of precursor coagulation species}

The previous section investigates how varying the platelet  packing density impacts the overall accumulation of important coagulation and fibrin polymerization species, as well as how fibrin gel formation differs under varying flow conditions. To understand how fibrin gel formation depends on the plug microstructure, we investigate how individual species involved in coagulation and fibrin polymerization vary spatially for the different platelet packing densities.

     \begin{figure}[!h]          
    \includegraphics[width=\textwidth]{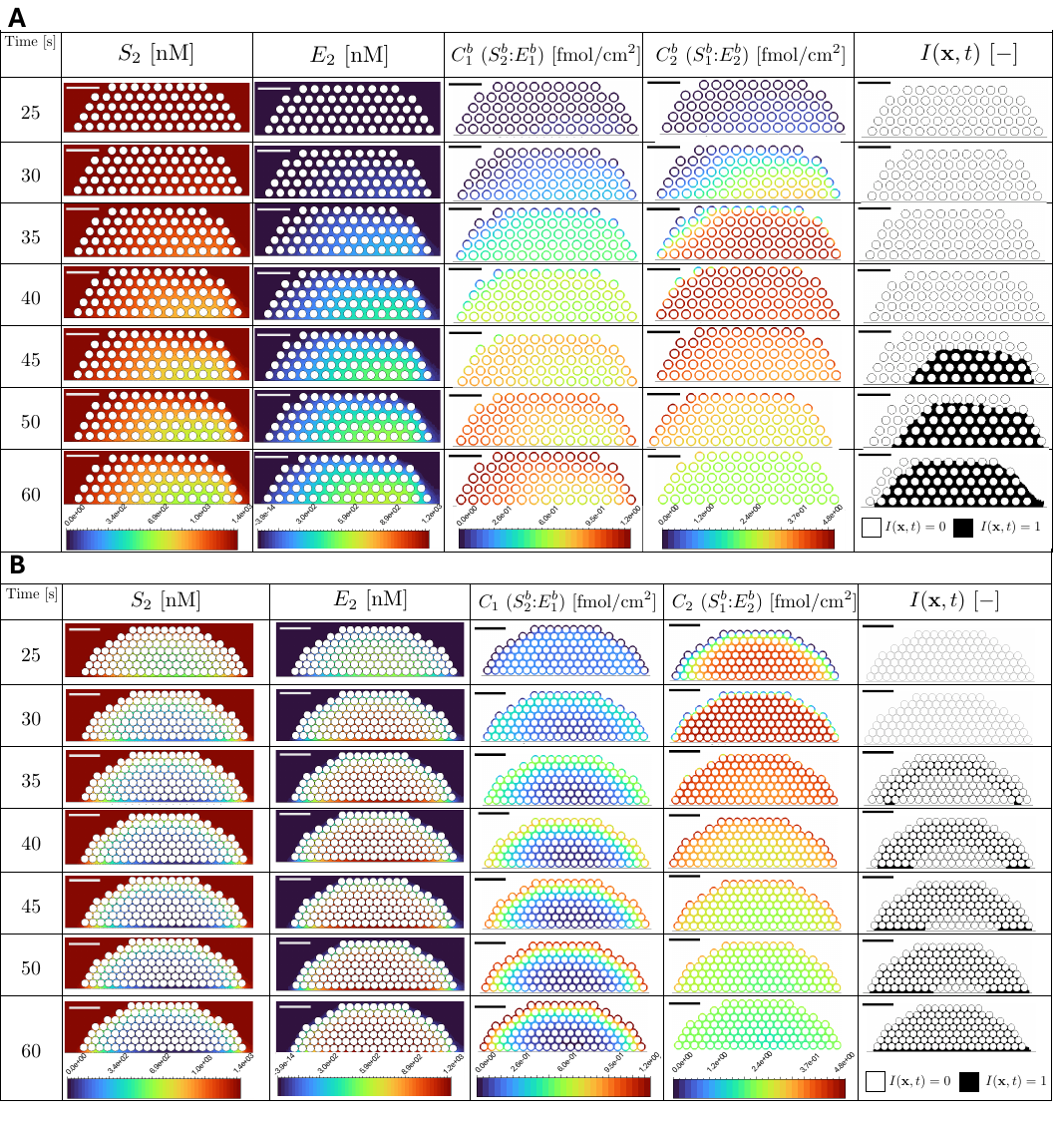}
        \caption{Concentrations and densities of coagulation species in the fluid and on platelet surfaces, respectively, for (A) the loose platelet plug and (B) the dense platelet plug. Each row corresponds to a different time value, the color bar for each column refers to the concentration or value of each variable, and the horizontal bar represents 10~\textmu m.}
        \label{fig:coag_2D} 
        \end{figure}

Local concentrations influence reaction rates, which impact the formation and motion of the fibrin gel front.  We show the concentrations of coagulation species, as well as the gel indicator variable, within the full 2D loose and dense platelet plugs in Fig.\ \ref{fig:coag_2D} starting at time 25 s. For the loose plug (Fig.\ \ref{fig:coag_2D}A), we see that the platelet-bound enzyme complex $C_2^b$ ($S_1^b:E_2^b$) accumulates first on the platelet surfaces in the near-wall downstream portion of the plug. Since the catalytic rate constant for the formation of $E_2$ from $C_1^b$ is more than an order of magnitude larger than that for the formation of $E_1$ from $C_2^b$, (see Table~\ref{tab:binding}), $C_2^b$ can accumulate on the surface. The $C_2^b$ complex then generates additional $E_1$ in the fluid which then forms more of the platelet-bound complex $C_1^b$ ($S_2^b:E_1^b$), which, in turn, produces more $E_2$. The accumulation of $E_2$ is greater in the near-wall region and is biased towards the downstream portion of the plug.

The concentrations of coagulation species for the dense plug results (\Cref{fig:coag_2D}B) are markedly different from these loose plug results. This is due to the early accumulation of $E_2$ and an early decrease of $S_2$ in the interior of the dense plug indicating that thrombin production occurs earlier than for the loose plug. The earlier production of thrombin corresponds to an earlier accumulation of $C_1^b$ (the complex responsible for the production of $E_2$) on the surface of the platelets within the center of the domain. Additionally, the thrombin and prothrombin concentrations within the dense platelet plug remain high and low, respectively, throughout the simulation, indicating that enzyme removal and substrate replenishment by transport are small; this is not the case for the loose plug.

\subsection*{Temporal dynamics of fibrin polymerization species depend on spatial location and platelet packing density   }

 \begin{figure}[!h]      \includegraphics[width=\textwidth]{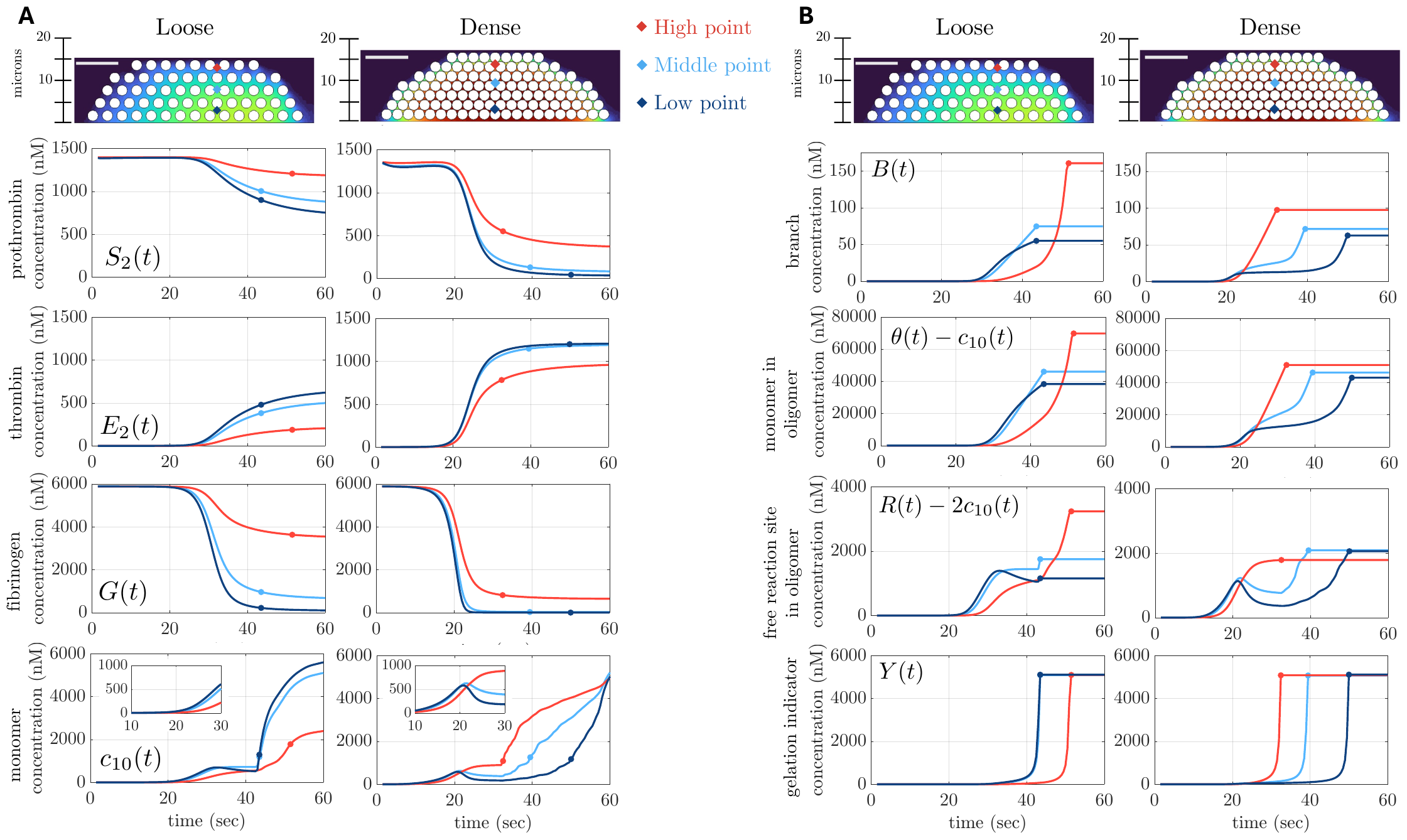}
        \caption{Concentrations over time of (A) coagulation species, and fibrin polymerization species, and (B) structural quantities for the (left column) loose and the (right column) dense platelet plugs. Line color refers to three spatial locations within the platelet plug: (red) high, (light blue) middle, (dark blue) low. Each dot corresponds to the gel time at that spatial location. Horizontal bar corresponds to 10 \textmu m.}
        \label{fig:structure_coag_2d}
        \end{figure}

We next seek to understand how the nonuniform spatial accumulation of thrombin impacts fibrin polymerization. To allow examination of the time-courses at different spatial locations with more temporal resolution than is possible using the spatial plots of \Cref{fig:coag_2D}, three spatial locations (high, middle, low) are chosen within the loose and dense platelet plugs that indicates the center of the region in which thrombin accumulates within the aggregate (top row in \Cref{fig:structure_coag_2d}). In \Cref{fig:structure_coag_2d}, we look at how various important concentrations change over time at these points. On each curve in this figure, a circle indicates the time of gelation at the corresponding location. 

In \Cref{fig:structure_coag_2d}A, the precursor species prothrombin ($S_2$) and fibrinogen ($G$) at all three spatial locations have lower concentrations later in time for the dense plug than for the loose one. In the dense plug, prothrombin is quickly converted to thrombin and is not replenished due to higher concentration of platelet reaction sites and lower effectiveness of transport. Consequently, thrombin accumulates and is not depleted due to flow, quickly converting fibrinogen within the plug interior which results in a rise in the fibrin monomer concentration, $c_{10}$ (seen in \Cref{fig:structure_coag_2d}A inset). Due to rapid depletion of fibrinogen within the interior of the dense plug, $c_{10}$ then declines at the lower two spatial points, whereas $c_{10}$ continues to increase at the high point due to more fibrinogen availability. For the loose plug configuration in \Cref{fig:structure_coag_2d}A, lower platelet reaction site concentration and more replenishment due to flow leads to less depletion of prothrombin and fibrinogen. Moreover, replenishment by transport is more effective for the loose plug, especially at the top and middle points. This results in a later rise in $c_{10}$, which is initially limited to the lower two locations where $E_2$ is substantially higher than at the high location. 

At different times following the onset of gelation, $c_{10}$ rises more substantially at all locations in both plugs. This is a consequence of our assumption that fibrin can no longer oligomerize at a spatial location after gelation has occurred there. The rise in $c_{10}$ then can be due to the local production of fibrin monomers which are not incorporated into oligomers there or to transport of $c_{10}$ from nearby locations where  gelation has already occurred.  We discuss this further below. The circles in \Cref{fig:structure_coag_2d}A show that for both configurations, gelation occurs first near the location of early monomer buildup; this occurs close to the injured wall (low point) for the loose plug and near the luminal edge (high point) for the dense plug.

\Cref{fig:structure_coag_2d}B shows quantities related to fibrin polymerization. Since polymerization reactions cease at any spatial point once gelation occurs there, all polymerization quantities are constant after gelation.  After gelation occurs at a first location, movement of monomers produced there to nearby locations leads to increases in the values of $c_{10}$ (\Cref{fig:structure_coag_2d}A) and contributes to polymerization at those locations. Values of $R$, $B$, and $\theta$ (\Cref{fig:structure_coag_2d}B) at these nearby locations increase and this potentially results in the spread of gelation to those adjacent areas. For the loose plug, gelation proceeds from near simultaneous occurrence at the lower and middle points to later occurrence at the high point. In contrast, for the dense plug, gelation starts and the high point and then proceeds sequentially to the middle and low points. Interestingly, \Cref{fig:structure_coag_2d}B shows that, at gel time, the concentrations of monomers in multimers ($\theta - c_{10}$), branches $B$, and free reaction site
concentration in multimers $R-2c_{10}$ are much higher for the high point in the loose plug than for the high point in the dense plug. In contrast, the loose and dense plugs have comparable values of these concentrations at gel time for the low and middle points.  We discuss the causes of this difference below.
        
We only show temporal results for three spatial locations. In the \nameref{sec:2D_trans_fibrin}, the full 2D concentration fields for fibrin polymerization species are shown in Fig S6. There, we see that the fibrin polymerization species are distributed at either the upstream or downstream end of the platelet plug, depending on the plug configuration. We have also focused only on results for shear rate 1000 s$^{-1}$. In \nameref{sec:flow_studies}, we compare results for selected spatial locations in the loose and dense plugs obtained in simulations with shear rate 1000 s$^{-1}$ with results obtained in simulations without flow. Figures S8, S9, S10, in \nameref{sec:flow_studies} show that without flow fibrin monomer supply and consumption rates, as well as fibrin species concentrations are lower for both plug densities. As a result, gelation occurred within 60 seconds at all three locations for the dense plug with flow, at only the top and middle points for the dense plug without flow, and at none of the locations for the loose plug without flow.

\subsection*{Fibrin monomer concentration and transport influence gelation location and motion}

\begin{figure}[!h]  \includegraphics[width=\textwidth]{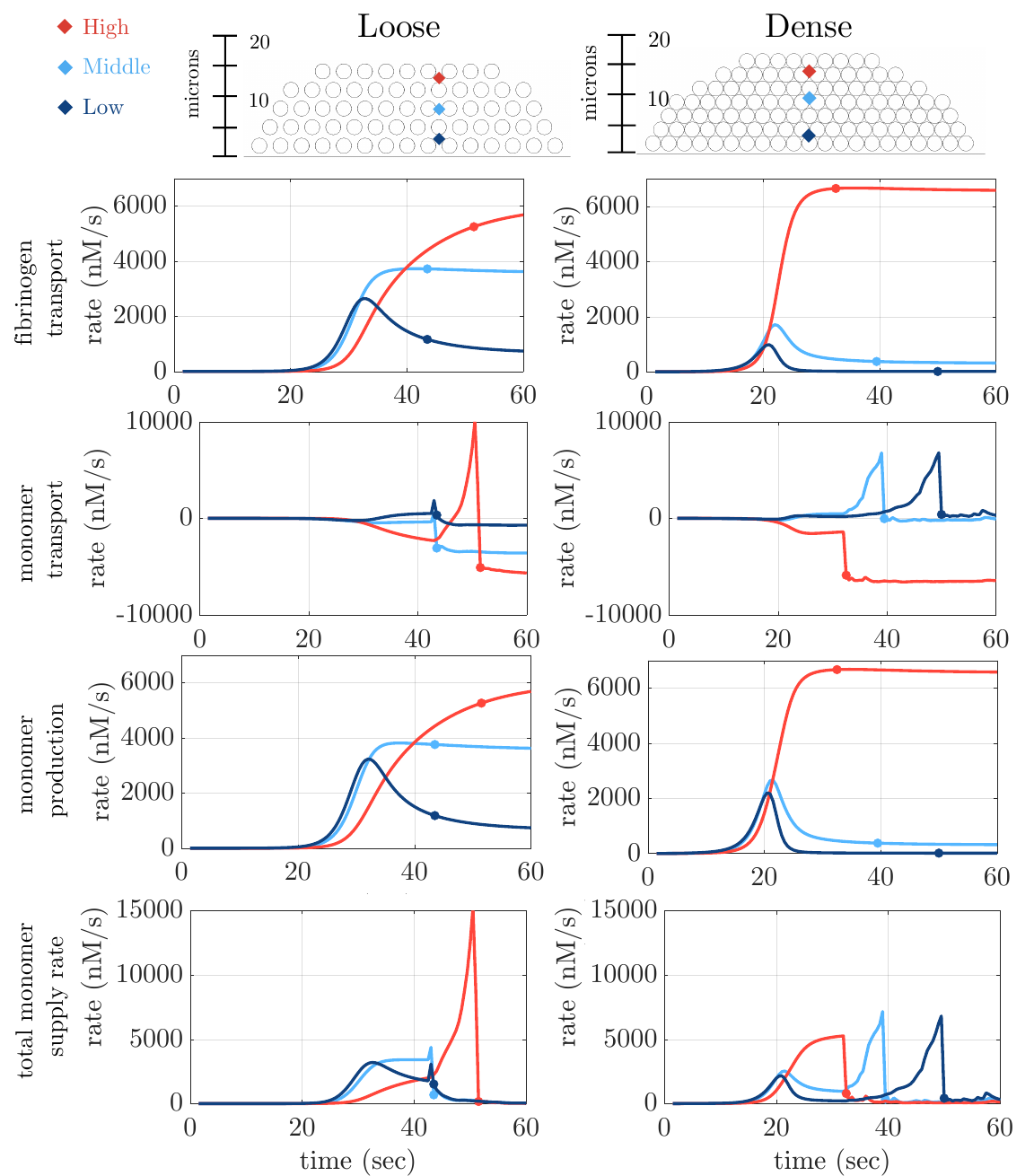}
 \caption{Rates over time (from top to bottom row) of fibrinogen transport ($-\mathbf{u} \cdot\nabla G + D\Delta G$), monomer transport ($-\mathbf{u} \cdot\nabla c_{10} + D\Delta c_{10}$), monomer production ($k_{\text{cat}}\frac{E_2G}{K_m + G}$) and total monomer supply rate ($k_{\text{cat}}\frac{E_2G}{K_m + G} + -\mathbf{u} \cdot\nabla c_{10} + D\Delta c_{10}$). Line color refers to three spatial locations within the platelet plug: (red) high, (light blue) middle, (dark blue) low. Each dot corresponds to the gel time at that spatial location.} 
\label{fig:transport_2d}
\end{figure}

The results in Figures ~\ref{fig:coag_2D} and
\ref{fig:structure_coag_2d} indicate that platelet packing density impacts the spatial location of gelation. In particular, the regions where $c_{10}$ accumulates early within the platelet plug correlate with where the gelation indicator $I$ first becomes nonzero. To understand this behavior, we investigate the mechanisms by which $c_{10}$ is produced in a given spatial region. Figure~\ref{fig:transport_2d} shows the impact of platelet packing density on the rate of transport of fibrinogen, $-\mathbf{u} \cdot\nabla G + D\Delta G$, and fibrin monomers, $-\mathbf{u} \cdot\nabla c_{10} + D\Delta c_{10}$, as well as the rate of production of monomer and the total monomer supply rate. We define the fibrin monomer production rate as $k_{\text{cat}}E_2 \frac{G}{K_m + G}$ and the total monomer supply rate as the sum of the rates of monomer transport and monomer production.

In Fig.~\ref{fig:transport_2d}, fibrinogen transport strongly impacts the monomer production rates for both plugs, where positive values indicate transport to and negative values indicate transport away from the location. In both the dense and loose plugs, the rate of fibrinogen transport is almost identical to the rate of monomer production, indicating that the production of fibrin monomer is limited by the transport of fibrinogen. At the middle and low point in the dense plug, the maximum rate of transport of fibrinogen is almost an order of magnitude smaller than its maximum rate of transport in the interior of the loose plug. Based on the Peclet number in the dense plug (see Table~\ref{tab:plt_char}), transport in this case is primarily due to diffusion, while the Peclet number in the loose plug indicates diffusive and advective transport are both important. 

For both plugs, fibrin monomers leave the peripheral region near the high point (the transport rate there is negative) escaping into the bulk fluid and moving toward the plug's interior (the transport rates there are slightly positive).  At the high point in the dense plug and at the low and mid points for the loose plug, monomer transport becomes more negative once gelation occurs at those points, because monomers newly produced there are not incorporated into the oligomers after gelation.  As a consequence, we see that at the mid point in the dense plug, monomer transport becomes positive following gelation at the top point; similarly, monomer transport becomes positive at the top point in the loose plug following gelation within the plug interior.

The overall rate of supply of fibrin monomer at a specific location is the sum of the monomer production rate and monomer transport rates there.  We see that gelation occurs earliest at the location with the greatest sustained early rate of monomer supply.  For the loose plug, gelation occurs earliest and almost simultaneously at the mid and low points whereas for the dense plug, it occurs first at the high point.  Gelation occurs sooner for the dense plug than for the loose one due to the high rate of monomer production near the periphery of the dense plug; this is the consequence of the simultaneous presence of high thrombin and fibrinogen concentrations.  At the lower points in the dense plug, the monomer supply rate initially rises after thrombin is produced but quickly drops because of depletion of fibrinogen.  Progress towards gelation at these points is slow with a gradual increase in branch concentration $B$ and oligomeric mass $\theta - c_{10}$ as shown in \Cref{fig:structure_coag_2d}B.  But this progress becomes more rapid after gelation occurs at the top point because of transport of monomers to these points from the plug periphery. The transport and production rates in two dimensions shown \nameref{sec:2D_trans_fibrin} in Fig S7 also illustrate the movement of the gel front. In particular, we can clearly see the gel front there moving toward the periphery (from low to high) in the loose case and towards the injured wall (from high to low) in the dense case.

Due to our modeling assumption, the timing of the gelation is influenced by monomer transport from regions in which a gel has already formed. Recall that some assumption about post-gelation reactions is needed because the differential equations for $B$, $\theta$, and $R$ are not valid after gel time, so we assume that after gelation, further incorporation of monomers into oligomers at that location ceases.  We sought to obtain an estimate of the magnitude of the effect of this assumption on the timing of the subsequent gelation events. To do this, we modified a version of our non-spatial fibrin branching model \cite{fogelson2010toward} that incorporates as inputs the thrombin and fibrinogen concentration time courses from a particular location in our current simulations  to produce fibrin monomers. The resulting fibrin monomer concentration is therefore unaffected by fibrin transport from other locations and can be compared with results in \Cref{fig:structure_coag_2d}. Figure S2 in \nameref{S3_text} shows the gel times and time courses of $c_{10}$ and $R$ when this is done for the low point in the dense plug, and we find that gel times differ by only a few seconds. 

\section*{Discussion}
\label{sec:discussion}
The formation of a thrombus involves the interplay of platelet activation and aggregation, the production of thrombin through the coagulation reactions, and the subsequent formation of a fibrin mesh that stabilizes the growing platelet aggregate under flow conditions. Here, we present a mathematical framework that examines how physical characteristics of the platelet microstructure can impact the production of thrombin and fibrin polymerization. Our framework incorporates a reduced model of coagulation reactions, most of which occurs on the surfaces of platelets in a preformed plug or aggregate made up of discrete platelets.  The framework also includes a model of fibrin polymerization which allows for a branched fibrin mesh to form thus making gelation possible. Under flow, blood moves between the stationary platelets in the aggregate and therefore coagulation proteins and fibrin polymerization species can move within the aggregate by advection and diffusion.  To our knowledge, this is the first model where coagulation reactions are modeled on discrete platelet surfaces and subsequent fibrin polymerization in the surrounding space is tracked.

\subsection*{Fluid volume to platelet surface area influences initial location of fibrin gel}
By varying both the spacing between platelets in the plug and the flow conditions, we studied how they, along with coagulation reactions on discrete platelet surfaces, impact the concentrations of fluid-phase coagulation species and influence fibrin polymerization. The most surprising result we found in our simulations was that gelation occurred first at the periphery of the medium and dense plugs, whereas it occurred first near the vessel wall for the loose plug.  This spatial shift can be understood from the interplay between platelet plug density, thrombin generation, and fibrinogen transport. Dense plugs provide more platelet surface area but less fluid space between platelets, promoting rapid thrombin formation and leading to high local thrombin concentrations at some locations. This effect was mediated by the total amount of products generated on the platelet surface and was clearly seen when comparing the total amount and average concentration of substrates and products over the course of the simulation. 

The high concentration of thrombin in the dense plug impacts fibrin polymerization dynamics through the acceleration of fibrinogen consumption and monomer production. Due to extremely low fluid velocities within the dense plug, polymerization reactions occur at the plug periphery where fibrinogen replenishment is sustained, causing gelation to begin at the periphery and to propagate inward. In contrast, for the loose plug we see greater fibrinogen transport into the plug interior, especially at early time points. Although the lower platelet surface area leads to slower production of coagulation enzymes compared to that in the dense plug,  replenishment of fibrinogen supports persistent monomer generation deeper within the plug.  Consequently, gelation begins within the loose plug interior. Despite producing roughly twice as much monomer overall as the dense plug (\Cref{fig:total_conc}), the loose plug gels initially at a later time point and in a different spatial pattern than for the denser plugs, most likely due to washout in the loose plug.

\subsection*{Qualitative comparisons with experimental and computational literature}
Ideally, our predictions of where and when fibrin gelation occurs would be validated directly with experiments, whether from static in vitro clotting assays, microfluidic flow chambers, or in vivo imaging of vascular injury. However, a direct, quantitative comparison is not yet possible due to several simplifications in our modeling framework. Our simulations are 2D and use a preformed aggregate of stationary, rigid platelets with prescribed interplatelet gaps, so platelet deposition, aggregation, and contraction are not represented. We implement a reduced model of thrombin generation, which lacks important inhibitors, and thrombin binding to fibrin is neglected. Additionally, each experimental system differs from our computational setup in distinct ways. Static assays, often used in fibrin polymerization studies \cite{ryan1999structural}, lack flow-mediated transport and microfluidic devices use specific procoagulant surfaces. In vivo injury models involve 3D, dynamically evolving thrombi that are shaped by the vessel wall and other surrounding cells like red blood cells and immune cells. For these reasons, we do not attempt a direct comparison with any of these systems. Instead, we compare the qualitative trends predicted by our model with those observed experimentally.

Our simulations show increasingly hindered intraplug transport of fibrinogen as platelet packing density increases, causing fibrin formation to proceed sooner and faster near the plug's lumenal edge. This is consistent with in vivo experimental findings which observed limited plasma-protein accessibility to the dense platelet-core and observed a band-like zone around the dense core where plasma proteins could still penetrate \cite{stalker2013hierarchical}.  Other in vivo studies have shown that after vascular injury the early thrombus surface is platelet-rich, but as the thrombus stabilizes and contracts it becomes progressively covered by a fibrin network interspersed with platelet clusters \cite{kim2013fibrin,kamocka2010two}. Recent in vivo studies have shown that after vascular injury, fibrin accumulation occurs at the interface between the endothelial wall and the platelet aggregate \cite{ballard20244d}. Taken together, these observations suggest that our simulations agree qualitatively with several experimental results, and future work involves incorporating more physiologically relevant mechanisms into our framework to allow quantitative comparisons with these results.

While the flow rate and platelet packing density in our framework dictated where fibrin gelation first occurs, the timing of gelation was less affected by these attributes. These results are somewhat similar to results seen in microfluidic experiments which imaged whole-blood clot formation over time with shear rates of 300/s, 1000/s, and 3500/s \cite{eyisoylu2026arterial}. The authors tracked platelet deposition, fibrin formation, and von Willebrand factor deposition and saw comparable fibrin gel formation times for each shear rate and saw fibrin enclosing the platelet aggregate for high shear rate \cite{eyisoylu2026arterial}. Indeed, Fig S5 in \nameref{sec:2D_trans_fibrin} shows that across three platelet packing densities and three flow rates, the time of fibrin gel initiation differs only on the order of seconds. Furthermore, our simulations showed that with increasing shear rate, the total volume of gel decreased. This trend is consistent with experimental observations under flow. Swieringa et al. reported that at a relatively low venous shear rate (150 s$^{-1}$), the heights of platelet plugs were low but the plugs were surrounded by an extensive fibrin network that extended far outside the platelet aggregates, whereas at higher shear rates, fibrin outgrowth was more restricted \cite{swieringa2016platelet}. Likewise, Onasoga et al. demonstrated on TF–rich microspots that increasing the shear rate from 50 to 1000 s$^{-1}$ led to a significant reduction in fibrin gel height  \cite{onasoga2014thrombin}. In line with these studies, our results demonstrate that at higher shear the convective washout of thrombin and fibrin(ogen) factors limits the development of an extended ``fibrin tail" downstream of the plug. 

We can also compare our results with those found in other computational studies. The gap sizes used in our dense and medium plug models (0.1 \textmu m  and 0.5 \textmu m, respectively) fall within experimental gap size ranges found in \cite{Mirramezani2018_plateletpacking}, which  analyzed platelet aggregates formed over 10 minutes and reported interplatelet gap sizes ranging from 0.06 to 0.6 \textmu m. We also see that in the dense configuration, precursor species in our simulations were not replenished by flow and transport was dominated by diffusion. Other in silico experiments also indicate transport is due mainly to diffusion over physiologically relevant gap sizes and increasing the shear rate has little impact on velocity within the platelet aggregate   \cite{Mirramezani2018_plateletpacking}.   Our concentrations of thrombin were high compared to values reported in some studies of pure continuum models  \cite{montgomery2023clotfoam,leiderman2011grow,fogelson2012plateletcount,kuharsky2001surface} in which thrombin is distributed in the entire volume occupied by fluid and platelets. In our current work, the thrombin released from platelet surfaces into the fluid is confined to the fluid region between the platelets.  Some microfluidic experiments and computational results suggest that thrombin concentrations can be as high as 1000 nM within the thrombus core \cite{zhu2018contact,chen2019reduced}; however, the majority of thrombin produced there is likely bound to fibrin fibers, which our model does not account for.

\subsection*{Limitations and extensions}

While our model incorporates many essential components of the blood clotting process, it has several limitations. A significant limitation that hinders quantitative comparisons of in vivo and in vitro experimental results is that our computational framework is 2D, not 3D. Nevertheless, we have developed a mathematical framework that can simultaneously investigate platelet packing, transport, surface reactions, and fibrin polymerization under flow conditions. This framework allows for systematic investigations into how transport within the 2D clot is influenced by flow and platelet packing density, as well as quantify how gap volume affects surface-mediated reactions within the aggregate. An important future direction would be to extend our modeling framework to 3D. 

An additional assumption we made was to assume an initial, preformed platelet plug with varying yet uniform packing densities at the start of our simulations. This is not physiological, as platelet adhesion and cohesion are important dynamic components of hemostasis and as stated earlier, spatial heterogeneity of both platelet density and activation is seen in vivo \cite{stalker2013hierarchical,welsh2014systems, welsh2016systems, stalker2014systems,tomaiuolo2026volume}.  While we allowed the number of binding sites on our simulated platelets to vary in time, we did this uniformly and allowed all platelets to become fully procoagulant (see Eq.~\eqref{eq:plt_act}). Since experimental evidence suggests that only a subpopulation of densely-packed platelets within the core region of an aggregate can fully support coagulation \cite{stalker2014systems,stalker2013hierarchical,ballard20244d, tomaiuolo2026volume}, our model may overestimate the contribution of activated platelets.

Furthermore, we have implemented a reduced model of thrombin generation which accounts for the fact that thrombin production occurs on surfaces of activated platelets in the thrombus. Our main interest is capturing platelet surface-mediated thrombin generation, with thrombin produced to an extent and on time scales similar to those seen with a full model of coagulation \cite{fogelson2012plateletcount}. A more comprehensive description of coagulation can contain more than 130 differential equations and incorporates inhibitory pathways and multiple feedback loops \cite{ginsberg2025,leiderman2011grow}. Such a model can be used for a more precise dissection of the coagulation system in the presence of platelets and flow \cite{kuharsky2001surface,leiderman2016synergy}.

 In our fibrin polymerization model, gel times and, in turn, the gel structure, are sensitive to the value of trimolecular branching parameter, $k_b$. There is no known kinetic parameter for fibrin branching, so we have estimated this parameter with experimental data  \cite{ryan1999structural}. Incorporating more experimental data and using parameter estimation techniques to better quantify the branching rate remain future work. We also make simplifying assumptions on the transport of oligomers and monomers.  In order to derive the small set of ODEs for structural quantities of interest, one can either assume all oligomers diffuse with the same diffusion coefficient or a finitely-many can have differing diffusion coefficients. Here, we assume that only fibrinogen and fibrin monomers are transported in the fluid.  It is not difficult to extend transport to include a finite number of small oligomers as in \cite{fogelson2022development}, but we believe our current assumption is reasonable given that our interest in understanding situations where diffusion of large oligomers may be of little significance. If a blood clot forms in the extravascular space with other cells or if the platelet plug is sufficiently dense, the spatial gaps are likely to be small which will limit the importance of the diffusion of large oligomers.

Because the differential equations for the moments of the oligomer distribution (see Supplement \nameref{S3_text}) used to derive the differential equations for $R$, $\theta$, and $B$ are valid only until a gel is formed \cite{ziff1980kinetics,fogelson2010toward,nelson2023towards,fogelson2022development}, we were required to make a modeling assumption for how the dynamics at a particular spatial location ought to proceed after gelation occurs there.  A model in which oligomers could contribute to post-gelation thickening of fibers is not yet available, so we made the assumption that all polymerization reactions cease at a point in space when gelation first occurs there.  The consequence of this is that the only fate open to monomers produced there is to move to another location in the plug; this can alter the speed of polymerization reactions at that second location. In \nameref{sec:flow_studies}, we show evidence in Fig S10 that the magnitude of this effect is small, by comparing gel times in the simulations reported here with ones obtained using a non-spatial model (therefore having no transport). 

Beyond these limitations, our modeling framework can be used to study other important aspects of fibrin polymerization and coagulation under flow. Here, we chose to model preformed platelet plugs consisting of stationary, solid particles in a fluid domain without explicitly modeling physical connections between platelets. In reality, fibrinogen and fibrin polymers are important ligands for receptors on platelet surfaces like $\alpha$IIb$\beta$3 and GPVI, which are responsible for platelet aggregation and platelet activation, respectively \cite{litvinov2023blood,mangin2023glycoprotein}. Previous computational and experimental models have demonstrated that TF surface density exhibits a distinct threshold behavior for thrombus initiation, characterized by minimal clot formation at low TF levels and rapid thrombin generation above a critical TF concentration \cite{okorie2008determination,diamond2010tissue}. This model could investigate the impact of TF surface density and help understand the exact nature of the interaction between platelet plug microstructure, TF density, fibrin polymerization, and shear rate. Finally, our framework also makes predictions that can be tested directly as experimental imaging improves. In particular, our model predicts where fibrin gelation first appears within the platelet plug and how thrombin and fibrinogen concentrations vary across individual gaps within the platelet aggregate. Currently, existing measurements report fibrin and plasma-protein distributions in terms of fluorescence intensity over either the entire aggregate area or in regions such as the core or shell \cite{stalker2013hierarchical,welsh2014systems,swieringa2016platelet,eyisoylu2026arterial}. If advances in imaging resolve fibrin deposition and solute accumulation on the scale of individual platelets and our model is extended to include platelet aggregation, our model predictions have the potential to be tested experimentally.

\section*{Acknowledgments}
Research reported in this work was supported by the National Cancer Institute of the National Institutes of Health under award number U54CA272167 for ACN, and by the National Heart, Lung, and Blood Institute of the National Institutes of Health under award numbers R01HL151984 and R01HL157631 for ALF. The content is solely the responsibility of the authors and does not necessarily represent the official views of the National Institutes of Health. This work used the Dutch national e-infrastructure with the support of the SURF Cooperative using grant no. EINF-12214 for JMHC and FJHG. The authors acknowledge the EU funded project In Silico World, grant number 101016503 for JMHC and FJHG.

\newpage

\section*{Supporting information}
\subsection*{S1 Appendix.} 
\label{S1_text}
\paragraph*{Reduced coagulation model equations and details.} The reduced model describing the production of thrombin uses the following enzymatic reactions, which occur on the subendothelium      
\begin{equation}
E_0^b + S_1 \overset{k^{+}_{C_0}}{\underset{k^-_{C_0}}{\rightleftarrows}} C_0^b \xrightarrow{k^{\text{cat}}_{C_0}} E_0^b + E_1,
\end{equation}
or on platelet surfaces,
\begin{equation}
S_2^b + E_1^b \overset{k^{+,s}_{C_1}}{\underset{k^-_{C_1}}{\rightleftarrows}} C_1^b \xrightarrow{k^{\text{cat}}_{C_1}} E_2^b + E_1^b,
\end{equation}
\begin{equation}
S_1^b + E_2^b \overset{k^{+,s}_{C_2}}{\underset{k^-_{C_2}}{\rightleftarrows}} C_2^b \xrightarrow{k^{\text{cat}}_{C_2}} E_2^b + E_1^b,
\end{equation}
\begin{equation}
P_1 + S_1 \overset{k^{+}_{S_1}}{\underset{k^-_{S_1}}{\rightleftarrows}} S_1^b,
\end{equation}
\begin{equation}
P_2 + S_2 \overset{k^{+}_{S_2}}{\underset{k^-_{S_2}}{\rightleftarrows}} S_2^b,
\end{equation}
\begin{equation}
P_1 + E_1 \overset{k^{+}_{E_1}}{\underset{k^-_{E_1}}{\rightleftarrows}} E_1^b,
\end{equation}
\begin{equation}
P_2 + E_2 \overset{k^{+}_{E_2}}{\underset{k^-_{E_2}}{\rightleftarrows}} E_2^b.
\end{equation}

\noindent
\textbf{Fluid-phase Species: $S_1$, $E_1$, $S_2$, $E_2$}

\begin{equation}
\frac{\partial S_1}{\partial t} = -\nabla \cdot \left( \mathbf{u} S_1 - D \nabla S_1 \right), 
\end{equation}
\text{with boundary conditions} 
\[
- D \, \hat{n} \cdot \nabla S_{1} \Big|_{\partial \Omega_{\text{inj}}} = -k_{C_0}^{+} S_1 E_0^b + k_{C_0}^- C_0^b,
\]
and
\[
- D \, \hat{n} \cdot \nabla S_{1} \Big|_{\partial \Omega_{\text{pla}}}
= -k_{S_{1}}^{+} \left\{ N_{1} - \left( S_{1}^{b} + E_{1}^{b} + C_{1}^b + C_{2}^b \right) \right\} S_{1} 
+ k_{S_{1}}^{-} S_{1}^{b}.
\]

\begin{equation}
\frac{\partial S_2}{\partial t} = -\nabla \cdot \left( \mathbf{u} S_2 - D \nabla S_2 \right),
\end{equation}
\text{with boundary condition} 
\[
- D \, \hat{n} \cdot \nabla S_{2} \Big|_{\partial \Omega_{\text{pla}}} =
- k_{S_{2}}^{+} \left\{ N_{2}
- \left( S_{2}^{b} + E_{2}^{b} + C^b_{1} + C^b_{2} \right) \right\} S_{2} 
+ k_{S_{2}}^{-} S_{2}^{b}.
\]

\begin{equation}
\frac{\partial E_1}{\partial t} = -\nabla \cdot \left( \mathbf{u} E_1 - D \nabla E_1 \right), 
\end{equation}
with boundary conditions
\[
- D \hat{\mathbf{n}} \cdot \nabla E_1 \big|_{\partial \Omega_{\text{inj}}} = k_{C_0}^{\text{cat}} C_0^b, 
\]
and
\[
- D \hat{\mathbf{n}} \cdot \nabla E_1 \big|_{\partial \Omega_{\text{pla}}} =- k_{E_{1}}^{+} \left\{ N_{1} 
- \left( S_{1}^{b} + E_{1}^{b} + C_{1}^b + C_{2}^b \right) \right\} E_{1} 
+ k_{E_{1}}^{-} E_{1}^{b},
\]

\begin{equation}
\frac{\partial E_2}{\partial t} = -\nabla \cdot \left( \mathbf{u} E_2 - D \nabla E_2 \right) - k_{AT}  E_2,
\end{equation}
with boundary condition
\[
- D \hat{\mathbf{n}} \cdot \nabla E_2 \big|_{\partial \Omega_{\text{pla}}} = - k_{E_{2}}^{+} \left\{ N_{2} 
- \left( S_{2}^{b} + E_{2}^{b} + C_{1}^b + C_{2}^b \right) \right\} E_{2} 
+ k_{E_{2}}^{-} E_{2}^{b}.
\]

\noindent
\textbf{Subendothelium-bound Species: $E_0^b$, $C_0^b$}
\begin{align}
\frac{\partial E_0^b}{\partial t} &= -k_{C_0}^{+} S_1 E_0^b + \left( k_{C_0}^- + k_{C_0}^{\text{cat}} \right) C_0^b, \quad \text{on } \partial \Omega_{\text{inj}}  \\
\frac{\partial C_0^b}{\partial t} &= \phantom{-}k_{C_0}^{+} S_1 E_0^b - \left( k_{C_0}^- + k_{C_0}^{\text{cat}} \right) C_0^b, \quad \text{on } \partial \Omega_{\text{inj}}  
\end{align}

\noindent
\textbf{Platelet-bound Species: $S_1^b$, $E_1^b$, $S_2^b$, $E_2^b$, $C_1^b$, $C_2^b$} \\
\begin{align}
\frac{\partial S_1^b}{\partial t} &= k_{S_1}^{+} \left\{ N_1 - \left( S_1^b + E_1^b + C_1^b + C_2^b \right) \right\} S_1 - k_{S_1}^- S_1^b  - k_{C_2}^{+,s} S_1^b E_2^b + k_{C_2}^- C_2^b  \\
\frac{\partial S_2^b}{\partial t} &= k_{S_2}^{+} \left\{ N_2 - \left( S_2^b + E_2^b + C_1^b + C_2^b \right) \right\} S_2 - k_{S_2}^- S_2^b - k_{C_1}^{+,s} S_2^b E_1^b + k_{C_1}^- C_1^b  \\
\frac{\partial E_1^b}{\partial t} &= k_{E_1}^{+} \left\{ N_1 - \left( S_1^b + E_1^b + C_1^b + C_2^b \right) \right\} E_1 - k_{E_1}^- E_1^b \\
& - k_{C_1}^{+,s} S_2^b E_1^b + \left( k_{C_1}^- + k_{C_1}^{\text{cat}} \right) C_1^b + k_{C_2}^{\text{cat}} C_2^b \quad   \\
\frac{\partial E_2^b}{\partial t} &= k_{E_2}^{+} \left\{ N_2 - \left( S_2^b + E_2^b + C_1^b + C_2^b \right) \right\} E_2 - k_{E_2}^- E_2^b \\
& - k_{C_2}^{+,s} S_1^b E_2^b + \left( k_{C_2}^- + k_{C_2}^{\text{cat}} \right) C_2^b + k_{C_1}^{\text{cat}} C_1^b \quad   \\
\frac{\partial C_1^b}{\partial t} &= k_{C_1}^{+,s} S_2^b E_1^b - \left( k_{C_1}^- + k_{C_1}^{\text{cat}} \right) C_1^b,  \\
\frac{\partial C_2^b}{\partial t} &= k_{C_2}^{+,s} S_1^b E_2^b - \left( k_{C_2}^- + k_{C_2}^{\text{cat}} \right) C_2^b 
\end{align}

\newpage
\subsection*{S2 Appendix.} 
\paragraph*{Conversion from volumetric to surface rate rates.}
\label{S2_Text}
Our 2D computational framework introduces coagulation reactions on the 2D surfaces of discrete platelets, so platelet-bound species and rate constants are quantified as surface densities (amount per area). Experimental measurements and previous mathematical frameworks treated these species and reactions in the solution volume, so platelet-bound species are measured as volume concentrations and reaction rates are measured as volumetric rate constants. The following derivations show equations for both fluid-phase, volumetric species and equations for surface-bound species measured in amount per area. Here, we use variables and reaction rates with bars to denote surface-bound species, and no bars to represent volumetric species.\newline

{\noindent \bf{Derivation of platelet surface binding rates}.} First, we derive the rate for fluid-phase species to bind to the surfaces of platelets. As an example, we focus on the binding and unbinding of $S_1$ to the platelet receptor, $P_1$. To derive these rates, we first introduce an equation which describes all species and reaction rates as volumetric \cite{montgomery2023clotfoam}. Here, all reaction rates and species are be measured in $\frac{\mathrm{moles}}{\mathrm{dm}^3}$.

\begin{equation}
\label{eq:volum_bind}
-\frac{dS_1^b}{dt} = \frac{dS_1}{dt} = -k_{S_1}^{+}
\left(\underbrace{\frac{P^bN_1^{\text{plt}}}{N_A}}_{\substack{\text{total concentration} \\ \text{of binding sites}}} -  ~~\underbrace{\left( S_1^b + E_1^b + C_1^b + C_2^b \right) }_{\substack{\text{concentration of occupied} \\ \text{binding sites}}}\right) S_1
+ k_{S_1}^{-} S_1^b,
\end{equation}
where $P^b$ is the number of bound platelets per $\mathrm{dm}^3$ (liter) in a given volume, $V$, $N_1^{\text{plt}}$ is the number of $P_1$ receptors per platelet, and $N_A$ is $6.022\times 10^{23} \mathrm{~mol}^{-1}$ (Avogadro's constant). This implies that $\frac{dS_1}{dt}\times V$ would be $\frac{\mathrm{moles}}{\mathrm{sec}}$ of $S_1$ moving between fluid-phase and platelet-bound $S_1$. 

We now introduce a similar equation but with reaction rates and species measured as surface densities $\left(\frac{\mathrm{moles}}{\mathrm{dm}^2}\right)$ and denoted with a bar, so that

\begin{equation}
\label{eq:surf_bind}
-\frac{d\bar{S}_1^b}{dt}= -\bar{k}_{\bar{S}_1}^{+}
\left(\underbrace{\frac{N_1^{\text{plt}}}{N_A A_p}}_{\substack{\text{total density} \\ \text{of binding sites}}} -~~ \underbrace{\left( \bar{S}_1^b + \bar{E}_1^b + \bar{C}_1^b + \bar{C}_2^b \right) }_{\substack{\text{density of occupied} \\ \text{binding sites}}}\right) S_1
+ \bar{k}_{\bar{S}_1}^{-} \bar{S}_1^b,
\end{equation}
where $A_p= 4 \pi (0.5 \times P_{\text{diam}})^2$ is the characteristic platelet surface area measured in dm$^2$.  Note that since each term on the right-hand side must have units $\frac{\mathrm{moles}}{\mathrm{dm}^2~\mathrm{sec}}$, $\bar{k}_{S_1}^+$ must have units $\frac{1}{\frac{\mathrm{moles}}{\mathrm{dm}^3}~\mathrm{sec}}$.

Suppose that in volume $V$ there are $N^b$ bound platelets, so 
\begin{equation}
    P^b\, V = N^b,
\end{equation}
that each platelet has the same radius, and that all of these platelets see the same fluid-phase $S_1$. Then the rate at which the amount of $S_1$ moves between fluid-phase and platelet-bound $\left(\frac{\mathrm{moles}}{\mathrm{sec}}\right)$ in Equation~\eqref{eq:surf_bind} is
\begin{equation}
\label{eq:b}
-\left(N^b\, A_p\right)\frac{ d\bar{S}_1^b}{dt} =
-\bar{k}_{\bar{S}_1}^{+}N^b\, A_p\, 
\left(\frac{N_1^{\text{plt}}}{N_A\, A_p}
- \left( \bar{S}_1^b + \bar{E}_1^b + \bar{C}_1^b + \bar{C}_2^b\right) \right)S_1
+ \bar{k}_{\bar{S}_1}^{-}\left(N^b\, A_p\right)\bar{S}_1^b,
\end{equation}
and the rate of the amount of $S_1$ moving between fluid-phase and platelet-bound $\left(\frac{\mathrm{moles}}{\mathrm{sec}}\right)$ in Equation~\eqref{eq:volum_bind} is
\begin{equation}
\label{eq:volum_bind_trans}
-V\frac{dS_1^b}{dt} =  -k_{S_1}^{+}V
\left(\frac{P^bN_1^{\text{plt}}}{N_A} - \left(S_1^b - E_1^b - C_1^b - C_2^b\right) \right) S_1
+ k_{S_1}^{-} VS_1^b.
\end{equation}
To have the same rate transfer for the unbinding of $S_1^b$, then 
\begin{equation}
     k_{S_1}^{-} \left(V S_1^b\right)
= \left(N^b\, A_p\, \bar{S}_1^b\right) \bar{k}_{\bar{S}_1}^{-} .
\end{equation}
Since $V S_1^b$ and $N^b A_p \bar{S}_1^b$ both represent the moles of $S_1^b$ in volume $V$, then we have that 
\begin{equation}
    k_{S_1}^{-} = \bar{k}_{\bar{S}_1}^{-}.
\end{equation}
Similarly, to have the same net transfer rate for the binding of $S_1$ then 

\begin{equation}
    k_{S_1}^{+}V
\left(\frac{P^bN_1^{\text{plt}}}{N_A} - \left(S_1^b + E_1^b + C_1^b + C_2^b\right)\right)  = \bar{k}_{\bar{S}_1}^{+}N^b\, A_p\, 
\left(\frac{N_1^{\text{plt}}}{N_A\, A_p}
- \left( \bar{S}_1^b + \bar{E}_1^b + \bar{C}_1^b + \bar{C}_2^b \right)\right).
\end{equation}
Again, the terms multiplied by $k_{S_1}^{+}$ and $\bar{k}_{\bar{S}_1}^{+}$, respectively, represent the number of moles of a given species in the volume $V$, so 
\begin{equation}
    k_{S_1}^{+} = \bar{k}_{\bar{S}_1}^{+}.
\end{equation}

{\noindent \bf{Derivation of enzymatic reaction rates on platelet surfaces}.} 
Next, we derive rate constants for reactions between platelet-bound species. As an example, we focus on the formation of the complex $C_1^b$. As before, we introduce an equation which describes all species and reaction rates in a volumetric sense and another equation with bar variables that describes the same species and reaction rates on a surface. These equations are  
\begin{equation}
\label{eq:vol_comp}
\frac{dC_1^b}{dt} = \underbrace{k_{C_1}^{+} S_2^b E_1^b}_{\text{concentrations}} - (k_{C_1}^{-} + k^{\text{cat}}_{C_1}) C_1^b,
\end{equation}
and
\begin{equation}
\label{eq:surf_comp}
\frac{d\bar{C}_1^b}{dt} = \underbrace{\bar{k}_{\bar{C}_1}^{+} \bar{S}_2^b \bar{E}_1^b}_{\text{surface densities}} - (\bar{k}_{\bar{C}_1}^{-} + \bar{k}^{\text{cat}}_{\bar{C}_1})\bar{C}_1^b,
\end{equation}
respectively. As before, suppose that in volume $V$ there are $N^b$ bound platelets with $P^b\, V = N^b$. We then look at the rate at which the amount of $C_1^b$ (in moles) changes over time, so then Equations~\eqref{eq:vol_comp} and \eqref{eq:surf_comp} become 
\begin{equation}
\label{eq:vol_mole}
V\frac{dC_1^b}{dt} = k_{C_1}^{+} V\,S_2^b E_1^b - (k_{C_1}^{-} + k^{\text{cat}}_{C_1}) V \,C_1^b,
\end{equation}
and
\begin{equation}
\label{eq:surf_mole}
N^b\,A_p\,\frac{d\bar{C}_1^b}{dt} = \bar{k}_{\bar{C}_1}^{+} \,\bar{S}_2^b \,\bar{E}_1^b \,N^b\,A_p- (\bar{k}_{\bar{C}_1}^{-} + \bar{k}^{\text{cat}}_{\bar{C}_1}) \bar{C}_1^b\, N^b\,A_p,
\end{equation}
respectively. In order to have the same rate transfer for the formation of $C_1^b$, then 
\begin{equation}
\label{eq:balance}
    k_{C_1}^{+} V\,S_2^b = \bar{k}_{\bar{C}_1}^{+} \bar{S}_2^b \, \bar{E}_1^b\, N^b\,A_p. 
\end{equation}
Note that the molar amount of $E_1^b$, $S_2^b$ in a volume $V$ can be written as 
\begin{equation}
E_1^b V = N^b\, A_p\, \bar{E}_1^b, \qquad S_2^b V = N^b\, A_p\, \bar{S}_2^b,
\end{equation}
respectively. Rewriting the terms in Equation~\eqref{eq:balance}, we find that 
\begin{equation}
\frac{k_{C_1}^{+}}{V}
\left(S_2^b \,V\right)\left(E_1^b\, V\right)
= \bar{k}_{\bar{C}_1}^{+}
\frac{\left(N^b\,A_p\, \bar{S}_2^b\right)\left(N^b\,A_p\, \bar{E}_1^b\right)}{N^b\, A_p}.
\end{equation}
This equation necessitates that 
\begin{equation}
\frac{k_{C_1}^{+}}{V} = \frac{\bar{k}_{\bar{C}_1}^{+}}{N^b\, A_p},
\end{equation}
Let $k_{C_1}^{+,s} = \bar{k}_{\bar{C}_1}^+$ and let $k_{C_1}^{+,v}= k_{C_1}^+$, then we have that 
\begin{equation}
    k^{+,s}_{C_1}=\frac{N^b}{V}\, k^{+,v}_{C_1} \, A_p, 
\end{equation}
and since $N^b = P^b \, V$, then 
\begin{equation}
    k^{+,s}_{C_1}=P^b\, k^{+,v}_{C_1} \, A_p.
\end{equation}
Similarly, to have the same rate transfer for the disassociation and the catalytic conversion of $C_1^b$, 
\begin{equation}
    (k_{C_1}^{-} + k^{\text{cat}}_{C_1})\,V\,C_1^b = (\bar{k}_{\bar{C}_1}^{-} + \bar{k}^{\text{cat}}_{\bar{C}_1}) \,\bar{C}_1^b \, N^b\,A_p. 
\end{equation}
Since $V\,C_1^b$ and $\bar{C}_1^b \,N^b\,A_p $ both represent the molar amount of $C_1^b$ in a given volume, then 
\begin{equation}
    \bar{k}_{\bar{C}_1}^{-}  = {k}_{C_1}^{-}, \quad \bar{k}^{\text{cat}}_{\bar{C}_1} = k^{\text{cat}}_{C_1} . 
\end{equation}

To summarize, we convert the volumetric coagulation reaction rates, $k_*^{+,v}$, to surface coagulation reaction rates, $k_*^{+,s}$, using the following relationships:
\begin{equation}
    A_p=4  \pi  (0.5\times P_{\text{diam}})^2 
    \label{eq:ap}
\end{equation}
\begin{equation}
    k^{+,s}_{C_1}=P^b \times k^{+,v}_{C_1} \times A_p
\end{equation}
\begin{equation}
    k^{+,s}_{{C_2}}=P^b \times k^{+,v}_{C_2} \times A_p
\end{equation}
\begin{equation}
    N_1^{\text{plt}}=2700, \quad N_2^{\text{plt}} = 2000,
\end{equation}
\begin{equation}
    N_1=N_1^{\text{plt}}/(N_A\times A_p)
\end{equation}
\begin{equation}
    N_2=N_2^{\text{plt}}/(N_A\times A_p)
    \label{eq:n2}
\end{equation}
with $P^b=5.51 \times 10^{13}$ platelet/dm$^3$, a characteristic number density of platelets in the aggregate, $A_p=19.1$ \textmu m$^2$ a characteristic platelet surface area, and $N_A=6.022\times 10^{23}~\mathrm{mol}^{-1}$ (Avogadro's constant). Here, $k^{+,v}_{C_1}$, $k^{+,v}_{C_2}$ are the volumetric reaction rates for the given reactions from \cite{montgomery2023clotfoam}.

For the subendothelium surface, we have binding and unbinding of fluid-phase species to $E_0$, but we have no reactions between pairs of subendothelium-bound species. Hence, there is no need to change additional subendothelium reaction rates to surface densities.

\newpage
\subsection*{S3 Appendix.}
\label{S3_text}
\paragraph*{Fibrin polymerization model details.} The model of fibrin polymerization is extended from \cite{fogelson2010toward}, where here we allow fibrin monomers to advect and diffuse, and for fibrin monomer to be sourced from the enzymatic conversion of fibrinogen to fibrin by thrombin \cite{fogelson2022development}. A fibrin cluster can be uniquely described by two indices, $m$ and $b$, where $b$ describes the number of branch points in an oligomer and $m+2b$ is the total number of fibrin monomer in oligomer. Here, all species are are functions of spatial coordinates $\mathbf{x}$ and time $t$. The dynamic equations for $c_{mb}(\mathbf{x},t)$ is 
\begin{equation}
\begin{split}
\frac{\partial c_{mb}}{\partial t} &= \delta_{m1,}\delta_{b,0}\big( \nabla \cdot (D\nabla c_{10}) - \mathbf{u} \cdot \nabla c_{10}(\mathbf{x},t)\big)  + \delta_{m,1}\delta_{b,0}S_{mb}\\
&+ \underbrace{\frac{k_l}{2}\sum_{\substack{m_1 + m_2 = m \\ b_1+b_2 = b}} (b_1+2)(b_2+2)c_{m_1b_1}c_{m_2b_2} - k_l(b+2)c_{mb} R}_{\text{link formation}} \\
&+ \underbrace{\frac{k_{{b}}}{6} \sum_{\substack{m_1 + m_2 + m_3= m+2 \\ b_1+b_2+b_3 =b-1}}(b_1+2)(b_2+2)(b_3+2)c_{m_1b_1}c_{m_2b_2}c_{m_3b_3}-\frac{k_b}{2}(b+2)c_{mb}  R^2}_{\text{branch formation}},
\end{split}
\label{eq:pbe}
\end{equation}
where $\delta_{i,j}$ is the Kronecker delta. The Kronecker deltas multiplying the transport terms mean that only fibrin monomers, not dimers or larger oligomers move.  The Kronecker deltas multiplying the source term $S_{mb}$ means that only fibrin monomers are sourced.  The source of monomers is from conversion of fibrinogen by thrombin and occurs at the rate 
\begin{equation}
    S_{10} = k_{\text{cat}} E_2\frac{G}{K_m + G}. 
\end{equation}
Here, $G$ is the fibrinogen concentration, $E_2$ is the thrombin concentration, and $R$ is the concentration of free reaction sites. The equations for $c_{10}$, $G$, and $E_2$ are
\begin{equation}
\begin{aligned}
      \frac{\partial c_{10}}{\partial t}  &= -\mathbf{u} \cdot \nabla c_{10} + D\Delta c_{10} + k_{\text{cat}}E_2\frac{G}{K_m + G}-(2k_lR + k_bR^2)c_{10},
      \end{aligned}
\end{equation}
\begin{equation}
    \frac{\partial G}{\partial t} = -\mathbf{u} \cdot\nabla G + D\Delta G - k_{\text{cat}} E_2\frac{G}{K_m + G},
\end{equation}
and
\begin{equation}
    \frac{\partial E_2}{\partial t} = -\mathbf{u} \cdot\nabla E_2 + D\Delta E_2,
\end{equation}
respectively, and the free reaction site concentration, $R$ is defined as 
\begin{equation}
R = \sum_{m_1,b_1}(b_1+2)c_{m_1,b_1}. 
\end{equation}

To obtain differential equations for the moments that define the species of interest, we introduce the moment generating function
\begin{equation}
    g(\mathbf{x},t; y,z) = \sum_{m,b}y^mz^{b+2}c_{mb}(\mathbf{x},t).
\end{equation}
Substituting this function into Equation~\eqref{eq:pbe}, we obtain
\begin{equation}
\begin{split}
    \end{split}
    \label{eq:pde_g}
\end{equation}

Because only monomers move by advection and diffusion and only monomers are sourced,
\begin{equation}
    P(\mathbf{x},t;y,z) = yz^2\left(-\mathbf{u}\cdot\nabla c_{10} + \nabla \cdot (D\nabla c_{10}) + k_{\text{cat}} E_2\frac{G}{K_m + G}\right).
\end{equation}
To reformulate the problem to study moments of the system, we define moments 
\begin{equation}
    M_{jk} = \left.\frac{\partial^{j+k}g(\mathbf{x},t; y,z)}{\partial y^j \partial z^k}\right|_{y=1, z=1}.
    \label{eq:moment_func} 
\end{equation}
These moments can be related to the quantities of interest. For example,  the total concentration of free reaction sites, $R$, fibrin monomers in oligomer, $\theta$, and branches $B$ can expressed in terms of  first moments as
\begin{equation}
    R = \sum_{m,b} (b+2)c_{mb} = M_{01},
\end{equation}
\begin{equation}
    \theta = \sum_{m,b}(m+2b)c_{mb} = M_{10} + 2M_{01} - 4M_{00},
\end{equation}
\begin{equation}
    B  = \sum_{m,b} b c_{mb} =  M_{01} - 2M_{00}.
\end{equation} 

Using the PDE defined in Equation~\eqref{eq:pde_g} and the relationship in Equation~\eqref{eq:moment_func}, we can then derive PDEs that describe how $R$, $\theta$, and $B$ change. First, we define equations for moments $M_{00}$, $M_{10}$ and $M_{01} = R$, respectively: 
\begin{equation}
\begin{aligned}
        \frac{\partial M_{00}}{\partial t} &= -\frac{k_l}{2}R^2 - \frac{k_b}{3}R^3  -\mathbf{u}  \cdot \nabla c_{10} + D\Delta c_{10} + k_{\text{cat}}E_2\frac{G}{K_m+ G},\\
     \end{aligned}
 \end{equation}
 \begin{equation}
\begin{aligned}
        \frac{\partial M_{10}}{\partial t} &= - \frac{k_b}{3}R^3 -\mathbf{u}  \cdot \nabla c_{10} + D\Delta c_{10} + k_{\text{cat}}E_2\frac{G}{K_m+ G},\\
     \end{aligned}
 \end{equation}
\begin{equation}
\begin{aligned}
        \frac{\partial R}{\partial t} &= -k_lR^2 - \frac{k_b}{2}R^3 + 2\left(-\mathbf{u}  \cdot \nabla c_{10} + D\Delta c_{10} + k_{\text{cat}}E_2\frac{G}{K_m+ G}\right).\\
     \end{aligned}
 \end{equation}
From these equations, we can define equations for  branch point density and fibrin monomer in oligomer as
\begin{equation}
\frac{\partial B}{\partial t} = \frac{k_b}{6}R^3,
\end{equation}
and
\begin{equation}
\begin{aligned}
\frac{\partial \theta}{\partial t} &= -\mathbf{u} \cdot \nabla c_{10} + D\Delta c_{10} + k_{\text{cat}}E_2\frac{G}{K_m + G},
\end{aligned}
\end{equation}
respectively. 

Note that several equations above depend on the transport of fibrin monomer, $c_{10}$. For computational simplicity, we perform the following change of variables
\begin{equation}
Z = R - 2c_{10}, \qquad X = \theta - c_{10}, 
\end{equation}
where $Z$ corresponds to the concentration of reaction sites in oligomers and $X$ represents the concentration of monomers in oligomers. 

As described in \cite{ziff1980kinetics,fogelson2010toward,fogelson2022development}, the system of differential equations exhibits blow up in finite time, which we define as gelation. This can be interpreted as the emergence of an oligomer of infinite size, which is when the average oligomer size $A$ becomes infinite in finite time. This singularity can also be described as the time at which the second moment of the oligomer size distribution diverges. We  are motivated to choose this framework from in vitro, static experiments, which use ocular or rheological techniques to determine fibrin gelation time \cite{ryan1999structural,chernysh2008dynamic}. By utilizing a framework that mathematically defines gelation, we can then precisely quantify the time at which a fibrin gel emerges in our model. It is important to note that in our framework, we assume that fibrinogen, fibrin, and other larger oligomers do not occupy any space in our domain. Therefore, at each spatial point within our platelet aggregate, we track the concentration of each species and do not account for size exclusion effects or pore size. 

To determine the gelation time in our model, we can express $A$ as a linear combination of moments of order zero, one, and two. $A$ becomes infinite if and only if the specific second moment $M_{02}$ or the related quantity $Y = M_{02} - R$ does so.  The PDEs for $M_{02}$ and $Y$ are:
\begin{equation}
\begin{aligned}
 \frac{\partial M_{02}}{\partial t} &= k_l(M_{02}^2 - 2M_{02}R) + k_b(M_{02}^2 R- M_{02}R^2)\\
 &+ 2\left(-\mathbf{u}\cdot\nabla c_{10} + \nabla \cdot (D\nabla c_{10}) + k_{\text{cat}} E_2\frac{G}{K_m + G}\right),
 \end{aligned}
\end{equation}
and 
\begin{equation}
   \frac{\partial Y}{\partial t}= k_lY^2 + k_bR\left(\frac{R^2}{2} + RY + Y^2\right).
\end{equation}As stated in the manuscript, the differential equations are not valid after gelation, the time that $Y$ becomes unbounded at the spatial point in question. We define a binary gelation indicator variable, $I$, that indicates where gelation has occurred during the simulation. Once gelation occurs at a spatial point, $x$, we set $I = 1$ and all polymerization reactions stop so only advective and diffusive transport of fibrin monomer and fibrinogen can occur. We use the Kronecker delta in the polymerization terms to represent how gelation alters the dynamics of the following equations

\begin{equation}
\frac{\partial G}{\partial t} = -\mathbf{u}  \cdot \nabla G + D\Delta G-k_{\text{cat}}E_2\frac{G}{K_m+G},  
\end{equation}

\begin{equation}
    \begin{split}
\frac{\partial c_{10}}{\partial t}  &= -\mathbf{u}  \cdot \nabla c_{10} + D\Delta c_{10}
+ k_{\text{cat}} E_{2} \, \frac{G}{K_{m} + G} \\
&- \delta_{I,0}\left\{ 2k_{l}(Z + 2c_{10}) + k_{b}(Z + 2c_{10})^{2} \right\} c_{10},
\end{split}
\end{equation}

\begin{equation}
\begin{aligned}
\frac{\partial \theta}{\partial t} &= -\mathbf{u}\cdot\nabla c_{10} + D\Delta c_{10} + k_{\text{cat}} E_2\frac{G}{K_m + G},
\end{aligned}
\end{equation}

\begin{equation}
\frac{\partial Z}{\partial t} = \delta_{I,0}\left\{-k_l(Z + 2c_{10})^2 - \frac{k_b}{2}(Z + 2c_{10})^3 + 2\left(2k_1(Z + 2c_{10}) + k_b(Z + 2c_{10})^2\right)c_{10} \right\},
\end{equation}

\begin{equation}
\frac{\partial B}{\partial t} = \delta_{I,0}\left\{\frac{k_b}{6}(Z + 2c_{10})^3\right\} ,
\end{equation}

\begin{equation}
\frac{\partial X}{\partial t} =\delta_{I,0} \left\{ 2k_{l}(Z + 2c_{10}) + k_{b}(Z + 2c_{10})^{2} \right\} c_{10}m
\end{equation}

\begin{equation}
\frac{\partial Y}{\partial t} = \delta_{I,0}\left\{k_l Y^2 + k_b (Z + 2c_{10}) \left( \frac{(Z + 2c_{10})^2}{2} + (Z + 2c_{10})Y + Y^2 \right) \right\}.
\end{equation}

\begin{figure}[!h]
        \centering
         \renewcommand{\thefigure}{S1}
        \includegraphics[width=0.75\textwidth]{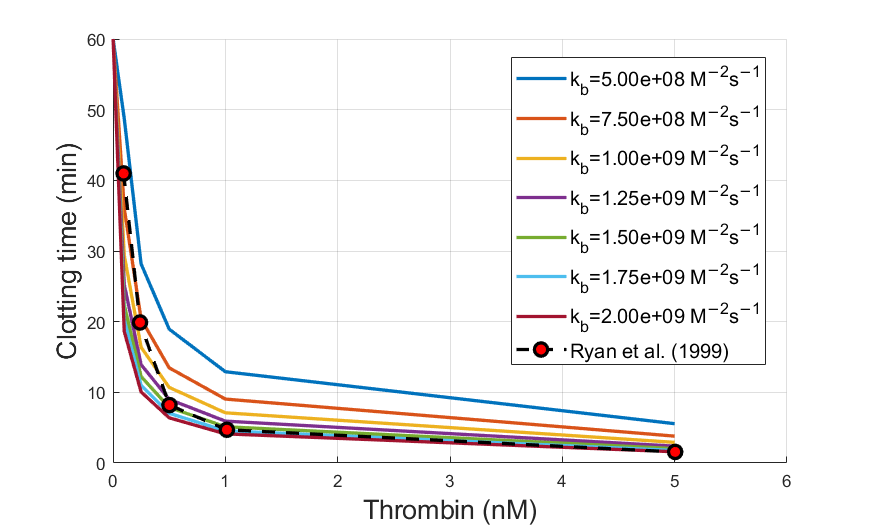}
        \caption{Gel time ($t_{gel}$) versus thrombin concentration as a function of different third-order rate constants, $k_b$, using the polymerization model found in \cite{fogelson2010toward}. }
                \label{S3_fig}
        \end{figure}

\paragraph*{Branching rate estimation from \cite{ryan1999structural}}
\label{sec:branch_est}
To estimate $k_b$ in Equation~\eqref{eq:pbe}, we simulate the original, zero-dimensional polymerization model in \cite{fogelson2010toward} with a Michaelis--Menten source term for fibrin monomer that depends on the thrombin and fibrinogen concentrations in Figure~\ref{S3_fig}. For different combinations of thrombin and fibrinogen concentrations and a fixed branching rate, $k_b$, the fibrin polymerization model will output the gel time, $t_{gel}$ which corresponds to the finite time blow up of $M_{02}$ and $Y$. We compare these computational gel times (color lines) to experimental gel times (red dots), where thrombin is varied and fibrinogen is fixed at 3 mg/mL \cite{ryan1999structural}. We find that $k_b = 1.5 \times 10^9$ M$^{-2}$s$^{-1}$ best approximates the behavior seen in experiments \cite{ryan1999structural}.  

\newpage

\paragraph*{Comparison of PDE model results with ODE model adapted from \cite{fogelson2010toward}}
\label{S10_fig}
Comparison of PDE results at the low point in the dense platelet plug with results from a zero-dimensional ODE model of fibrin polymerization in \cite{fogelson2010toward} in Figure~\ref{S10_fig}. With the modification, the source of fibrin monomer is fibrinogen converted to fibrin monomer by thrombin rather than a steady rate of supply as in \cite{fogelson2010toward}. In the modified ODE model, we use the time courses of fibrinogen $G$ and thrombin $E_2$ found in the PDE framework at that spatial location. 
           \begin{figure}[H]
        \centering
         \renewcommand{\thefigure}{S2}
        \begin{adjustwidth}{-0in}{0in}        \includegraphics[width=\textwidth]{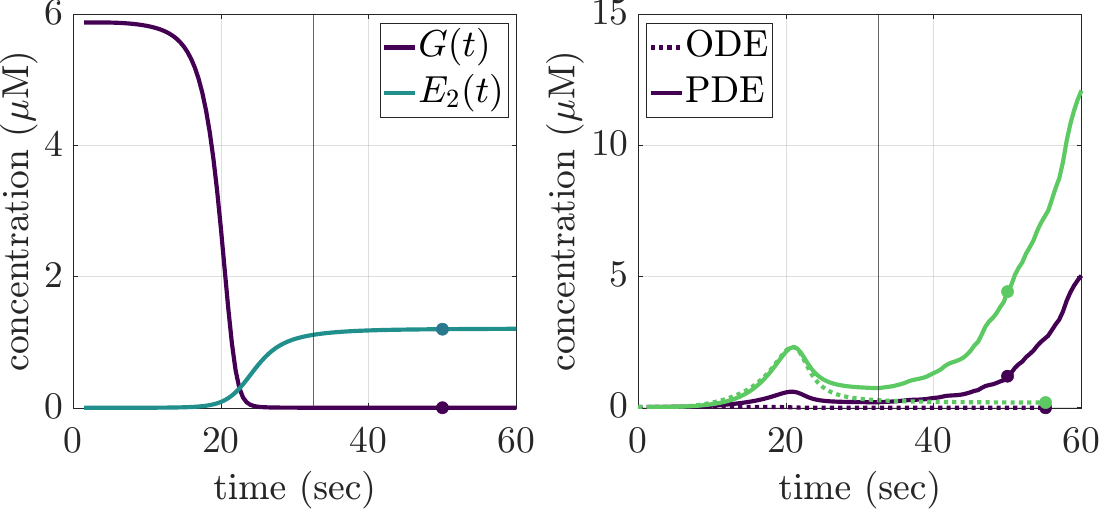}
        \caption{Left: Concentrations of fibrinogen and thrombin from the PDE model at the low point in the dense platelet configuration with $\dot\gamma = 1000$ s$^{-1}$. Right: Comparison of results from the full, PDE model (solid line) to zero-dimensional ODE model (dotted line) results from a modification of \cite{fogelson2010toward} with $G(t)$ and $E_2(t)$ from the PDE model as inputs. Green corresponds to free reaction site concentration, $R(t)$, and purple corresponds to fibrin monomer, $c_{10}(t)$. The dot in both panels refers to the gel time for the low spatial location for the PDE model, and vertical line refer to first gel time at the high point.}
        \label{S10_fig}
        \end{adjustwidth}
        \end{figure}

\newpage
\subsection*{S4 Appendix.}
\label{S4_text}
\paragraph*{Mesh and timestep convergence study} Preliminary simulations are performed with the loose plug at a shear rate of 1000 s$^{-1}$ to check for time
step and mesh independence. Three meshes
are generated for these tests, consisting of coarse, medium,
and fine refinement levels. The meshes have 11396, 29463, and 72051 computational cells, respectively. To assess temporal sensitivity, simulations were repeated on the fine mesh using time-step sizes ranging from 1 $\times$10$^{-2}$ to 5 $\times$10$^{-4}$ s, keeping all other parameters constant. To evaluate spatial sensitivity, simulations were repeated on the three mesh refinement levels, keeping all other parameters constant with a time-step size of 1 $\times$10$^{-3}$s. \begin{figure}[!h]
        \centering
          \renewcommand{\thefigure}{S3}
       \includegraphics[width=\textwidth]{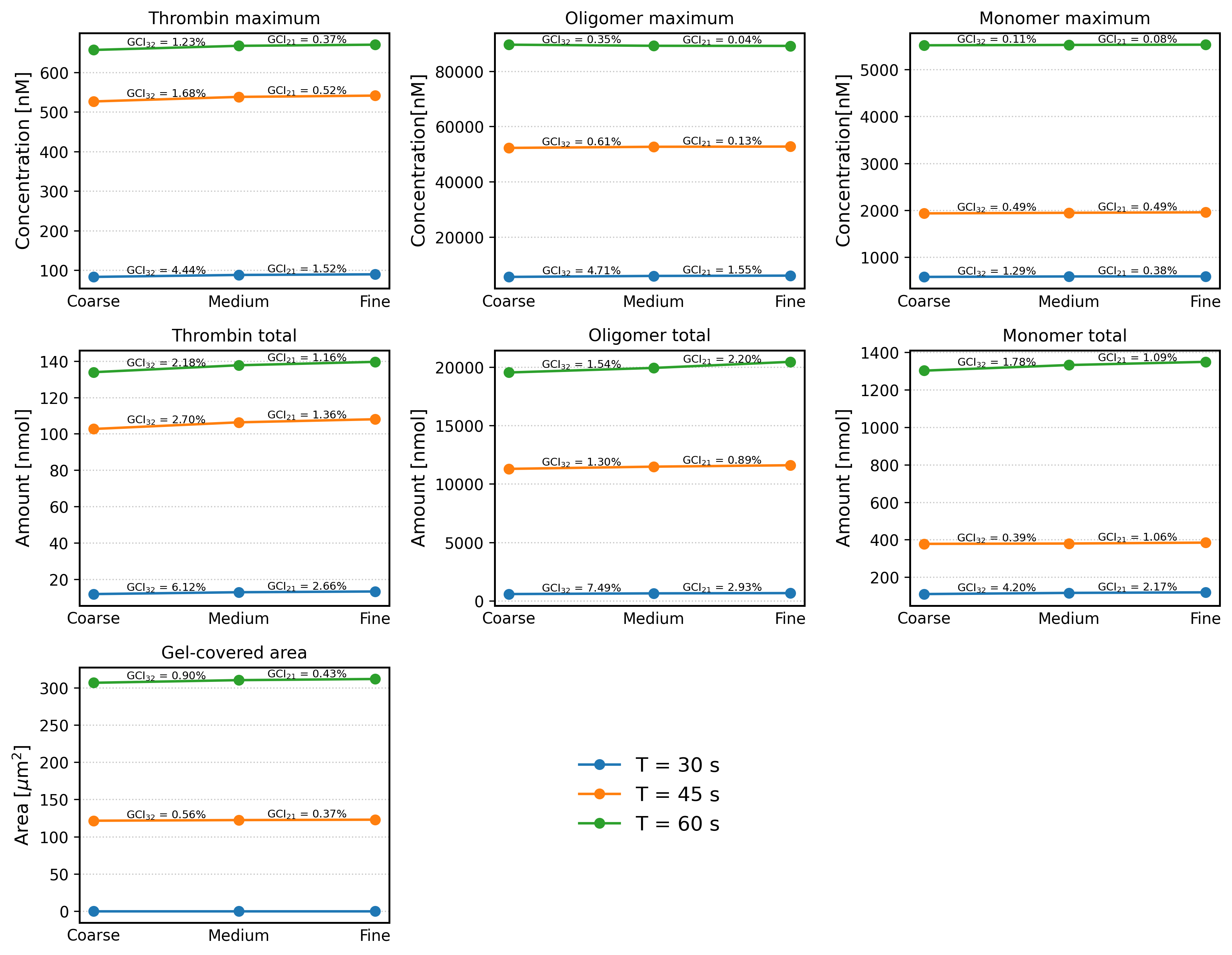}
        \caption{{\textbf{Spatial sensitivity of coagulation variables for different mesh resolutions.}} Spatial sensitivity of key coagulation variables at $\Delta t=10^{-3}$\,s for three mesh resolutions (coarse, medium, fine). Curves show results at $T=30,\,45,\,60$\,s with circular markers at each mesh configuration. The grid convergence index (GCI) is indicated for each panel, where the segment between Coarse$\!\rightarrow$Medium is annotated with $\mathrm{GCI}_{32}$ and the segment between Medium$\!\rightarrow$Fine with $\mathrm{GCI}_{21}$ (both in \%).
Panels are arranged by quantity: top row—maximum values (thrombin, monomers in oligomers, monomers); middle row—total domain‐integrated values (thrombin, monomers in oligomers, monomers); bottom row—gel‐covered area.}
        \label{S1_fig}
        \end{figure}
        
        The resulting spatiotemporal development of key coagulation species was compared across these simulations to verify time-step and mesh independence. The maximum concentration and total amount of thrombin ($E_2$), monomers ($c_{10}$), and monomers in oligomers ($\theta$), and the gel-covered area are measured at three time points: before gelation ($T=30$ s), directly after the gelation-onset ($T=45$ s), and after gelation finalization ($T=60$ s).   Following the method from \cite{roache_perspective_1994}, we determined the spatial discretization error as the grid convergence index. In this method, $Q_3,Q_2,Q_1$ are the quantities on the coarse, medium, and fine mesh, respectively, and the effective refinement ratio $r$ is estimated as $r_{23}\approx\sqrt{N_2/N_3}$, and $r_{21}\approx\sqrt{N_1/N_2}$, in which $N_i$ is the number of elements per mesh.  The relative errors are $e_{21}=\lvert (Q_1-Q_2)/ Q_1 \rvert$ and $e_{32}=\lvert (Q_2-Q_3)/ Q_2\rvert$. With safety factor $F_s=1.25$, and order p=2, the GCIs are

\[
\mathrm{GCI}_{21}=\frac{F_s\,e_{21}}{r_{21}^{p}-1},\qquad 
\mathrm{GCI}_{32}=\frac{F_s\,e_{32}}{r_{32}^{p}-1},
\]
We consider the mesh as spatially independent when $\mathrm{GCI}<5\%$. 

Temporal sensitivity on the fine mesh was quantified by successive percent changes between time-step levels $\Delta t_k\to\Delta t_{k+1}$:
\[
\Delta_k=\frac{\lvert Q(\Delta t_{k+1})-Q(\Delta t_{k})\rvert}{\lvert Q(\Delta t_{k+1})\rvert}\times 100\%
\]
We consider the time-step to be converged when $\Delta_k<3\%$.

\begin{figure}[!h]
        \centering
          \renewcommand{\thefigure}{S4}
        \includegraphics[width=\textwidth]{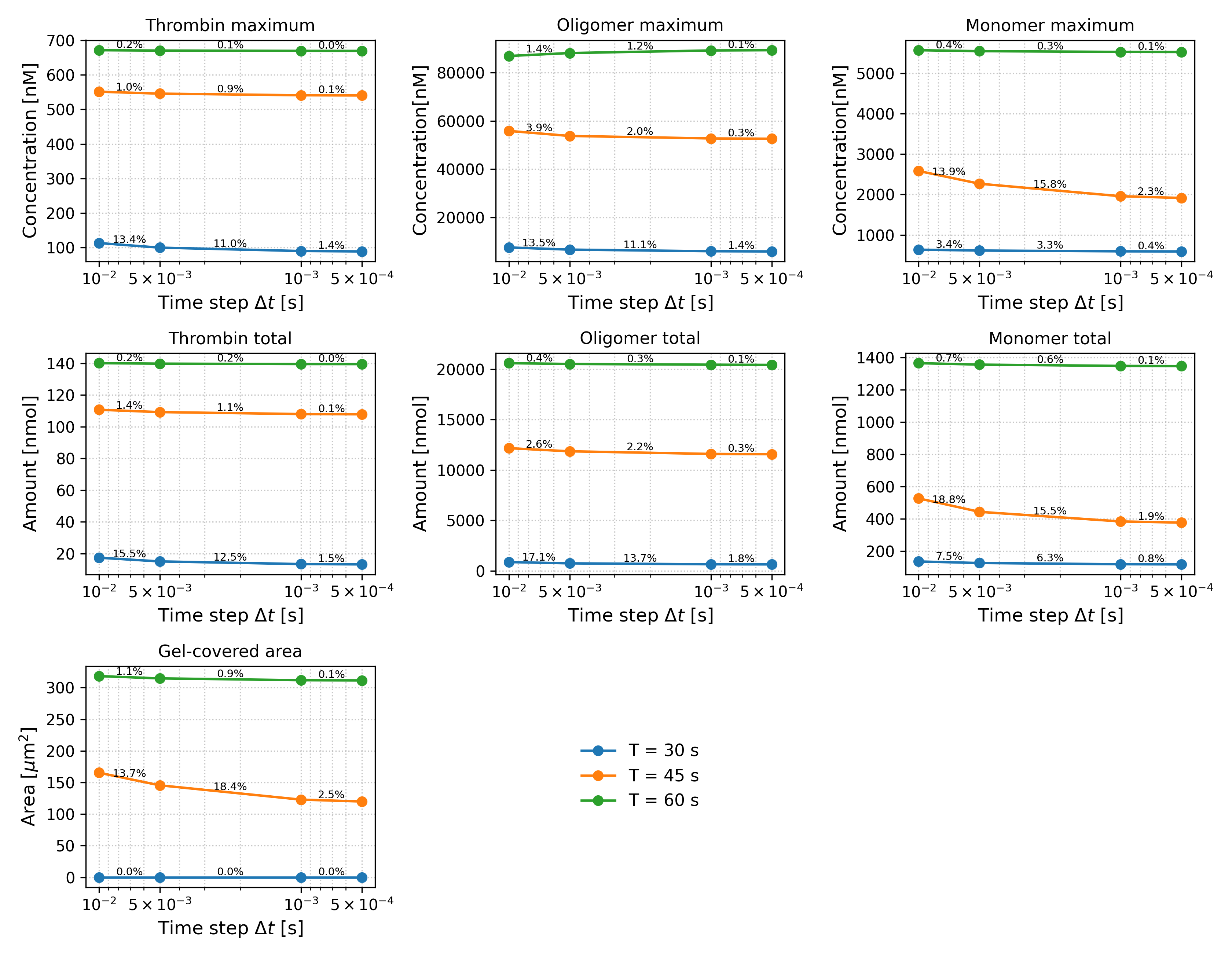}
        \caption{{\textbf{Temporal sensitivity of coagulation variables on the fine mesh at different time points.}}  Temporal sensitivity of key coagulation variables on the fine mesh at three analysis times ($T=30,\,45,\,60$\,s). Each subplot shows the variable versus the time step ($\Delta t=0.01\to0.005\to0.001\to0.0005\,$s), with $\Delta t$ on a logarithmic $x$-axis. Curves correspond to the four time points, with circular markers at the tested $\Delta t$ levels. The successive percent changes between adjacent $\Delta t$ levels are indicated. Panels are arranged by quantity: top row—maximum values (thrombin, monomers in oligomers, monomers); middle row—total domain‐integrated values (thrombin, monomers in oligomers, monomers); bottom row—gel‐covered area.}
        \label{S2_fig}
        \end{figure}

The three‐level spatial convergence study in Figure~\ref{S1_fig} shows that for all seven output metrics at $T=30\,$s, $T=45\,$s, and $T=60\,$s the fine‐grid GCI values are below 5\%. 
Therefore the medium mesh
is sufficient, reducing computational cost by $\sim$60\% compared to the fine mesh. The temporal convergence study in Figure~\ref{S2_fig} indicates that the change in the seven output metrics between $\Delta t=1\times10^{-3}\,$s and $\Delta t=5\times10^{-4}\,$s is below 3\% in all cases. Thus a time‐step of $\Delta t=1\times10^{-3}\,$s is sufficient for capturing the coagulation dynamics without significant temporal discretization error. Therefore, the medium mesh and a time step size of $1\times10^{-3}\,$s are adopted for the final simulations.\\

\clearpage

\subsection*{S5 Appendix.}
\label{sec:2D_trans_fibrin}
\begin{figure}[!h]
    \centering
          \renewcommand{\thefigure}{S5}
   \includegraphics[width=0.8\linewidth]{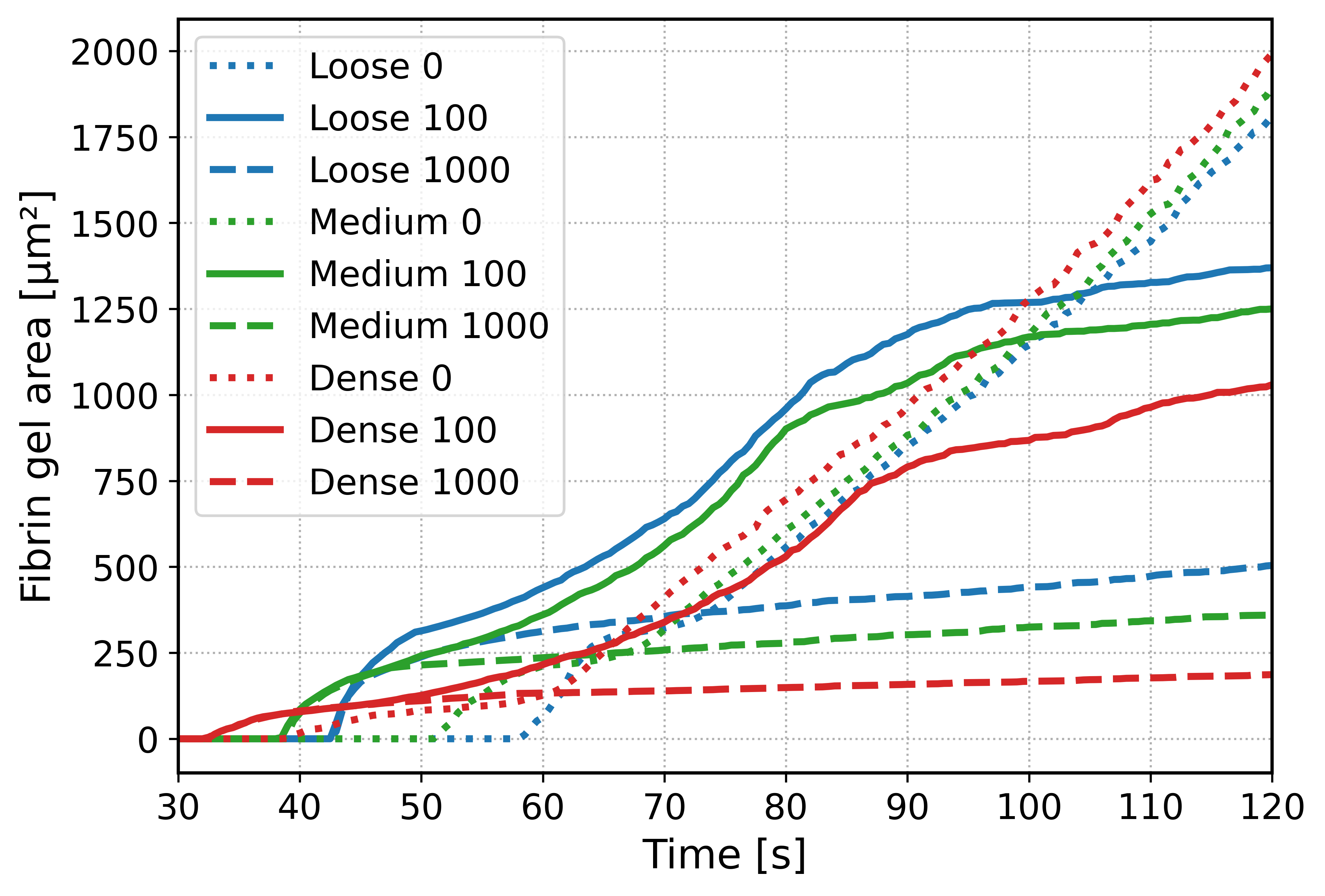}
   \caption{Time evolution of the fibrin gel area is shown for different platelet plug configurations (color) and different shear rates (line style) for up to 120 s. }
   \label{S4_fig}
\end{figure}

\begin{figure}[htbp]
 \centering
  \renewcommand{\thefigure}{S6}       \includegraphics[width=\textwidth]{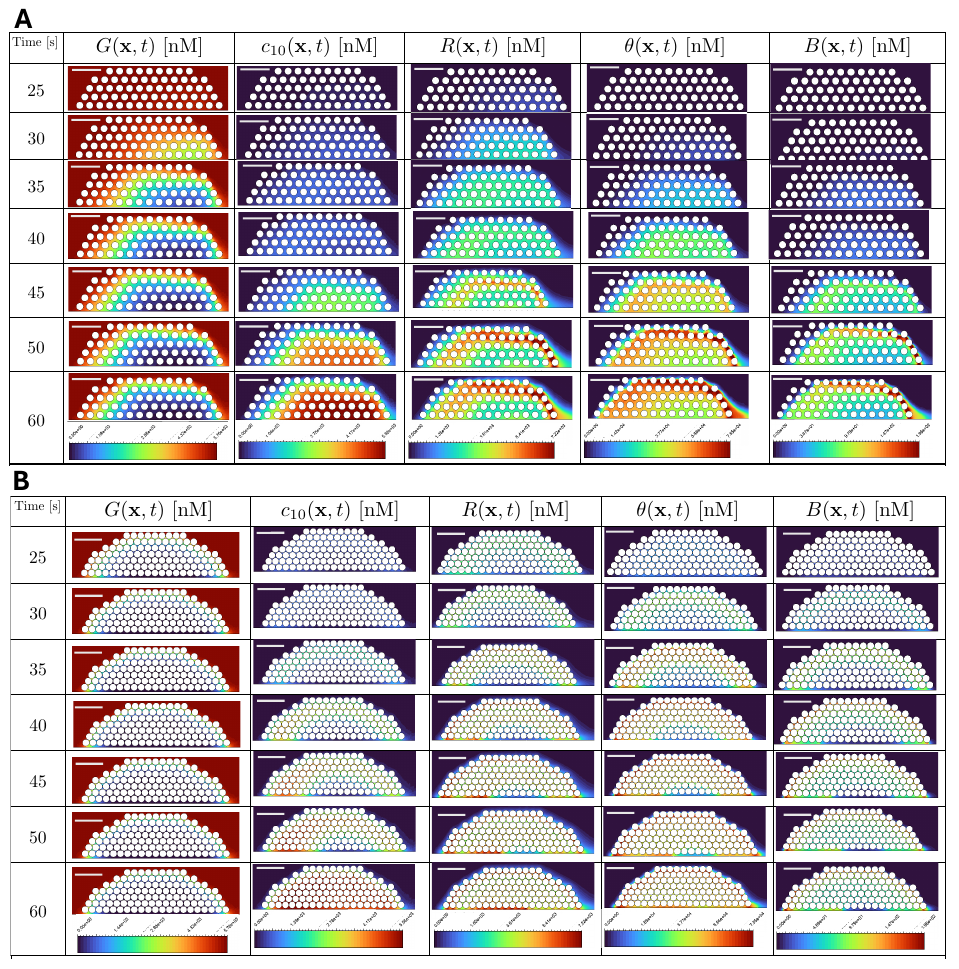}
 \caption{Concentrations of fibrin polymerization species in the fluid for (A) the loose platelet plug and (B) the dense platelet plug with shear rate $1000$ s$^{-1}$. Each row corresponds to a different time value, the color bar for each column refers to the concentration, and the horizontal bar represents 10~\textmu m. }
 \label{S5_fig}
        \end{figure}

\begin{figure}
  \centering
    \renewcommand{\thefigure}{S7}
 \begin{adjustwidth}{0in}{0in}        \includegraphics[width=\textwidth]{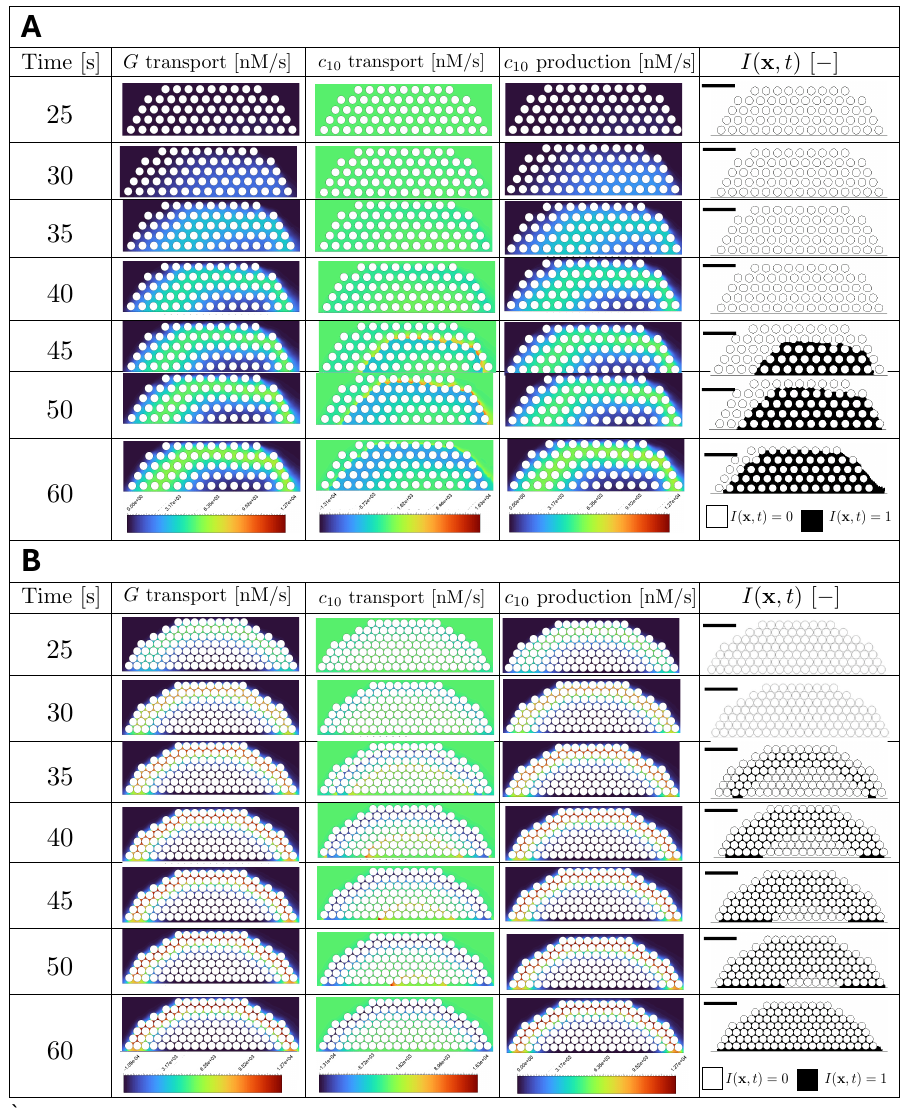}
    \end{adjustwidth}
    \caption{Transport (concentration per time) of fibrinogen $G(\mathbf{x},t)$ and transport and production (concentration per time) of fibrin monomer $c_{10}(\mathbf{x},t)$ over the domain around the platelet plug for (A) the loose platelet configuration and (B) the dense platelet configuration for shear rate $\dot \gamma = 1000$ s$^{-1}$. Positive and negative values indicate transport into and out of a point in the domain, respectively. Horizontal bar corresponds to 10 \textmu m.}
    \label{S6_fig}
        \end{figure}

\paragraph*{2D plots of gelation, transport, and fibrin polymerization species}
We show how the fibrin gel area changes as shear rate increases in loose, medium, and dense platelet configuration in \Cref{S4_fig}, as well as 2D plots of both fibrin polymerization species in \Cref{S5_fig} and transportation and production rates in \Cref{S6_fig}. Flow-mediated transport biases the distribution of fibrin polymerization species either toward the upstream or downstream ends of the plug, depending on the plug density.  In particular, \Cref{S5_fig}A shows that high concentrations of $R$, $\theta$, and $B$ are seen at the downstream (right-most) edge of the loose platelet plug, while \Cref{S5_fig}B shows that high concentrations of these variables are seen at the upstream (left-most) edge of the dense plug.

\clearpage

\subsection*{S6 Appendix.}
\label{sec:flow_studies}
\begin{figure}[H]
        \centering
         \renewcommand{\thefigure}{S8}
          \begin{adjustwidth}{-0.75in}{0in} 
         \includegraphics[width=1.2\textwidth]{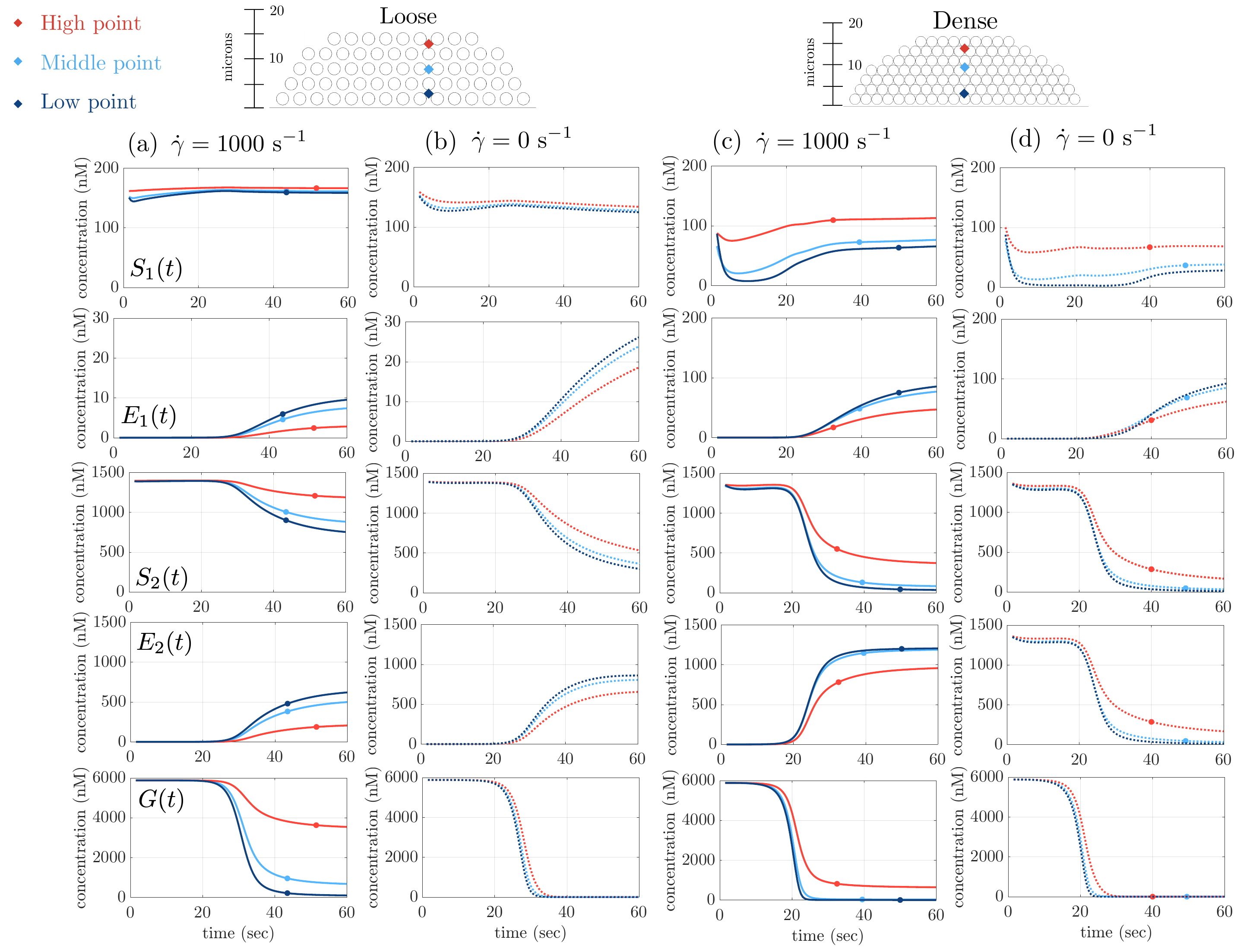}
 \caption{Time courses of coagulation species and fibrinogen concentrations for the loose platelet plug with (a) flow and (b) without flow and for the dense platelet plug with (c) flow and (d) without flow. Each color refers to a spatial location within the platelet plug: high point (red), middle point (cyan), and low point (blue). For each simulation, the dot corresponds to the time of gelation. Note that if there is no marker, gelation does not occur before 60 seconds. }
 \label{S7_fig}
    \end{adjustwidth}
        \end{figure}

\begin{figure}[H]
        \centering
         \renewcommand{\thefigure}{S9}   
             \begin{adjustwidth}{-0.75in}{0in} 
         \includegraphics[width=1.2\textwidth]{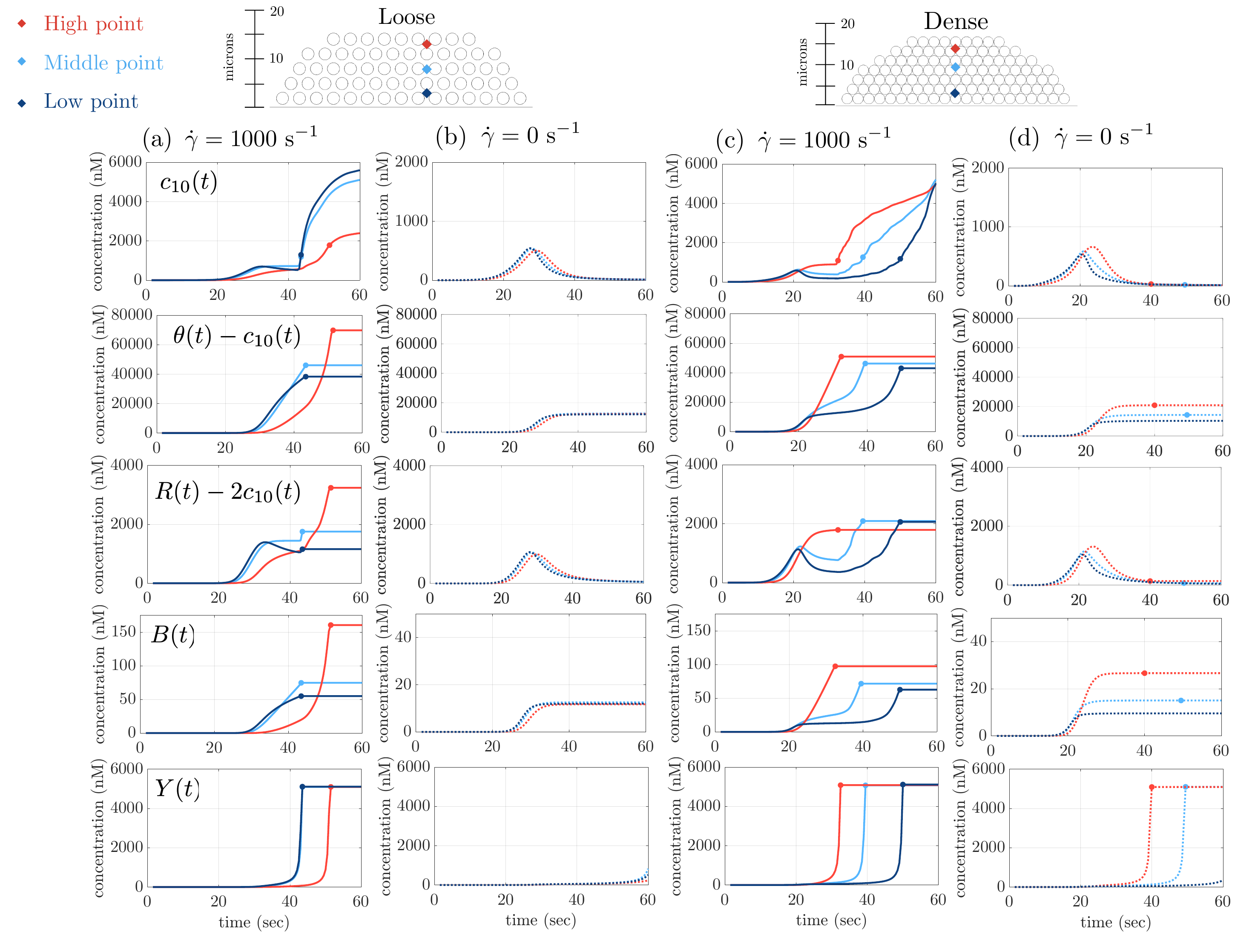}
 \caption{Time courses of fibrin polymerization species for the loose platelet plug with (a) flow and (b) without flow and for the dense platelet plug with (c) flow and (d) without flow. Each color refers to a spatial location within the platelet plug: high point (red), middle point (cyan), and low point (blue). For each simulation, the dot corresponds to the time of gelation. Note that if there is no marker, gelation does not occur before 60 seconds. }
 \label{S8_fig}
 \end{adjustwidth}
        \end{figure}

\begin{figure}[H]
        \centering
         \renewcommand{\thefigure}{S10}
             \begin{adjustwidth}{-0.75in}{0in} 
       \includegraphics[width=1.2\textwidth]{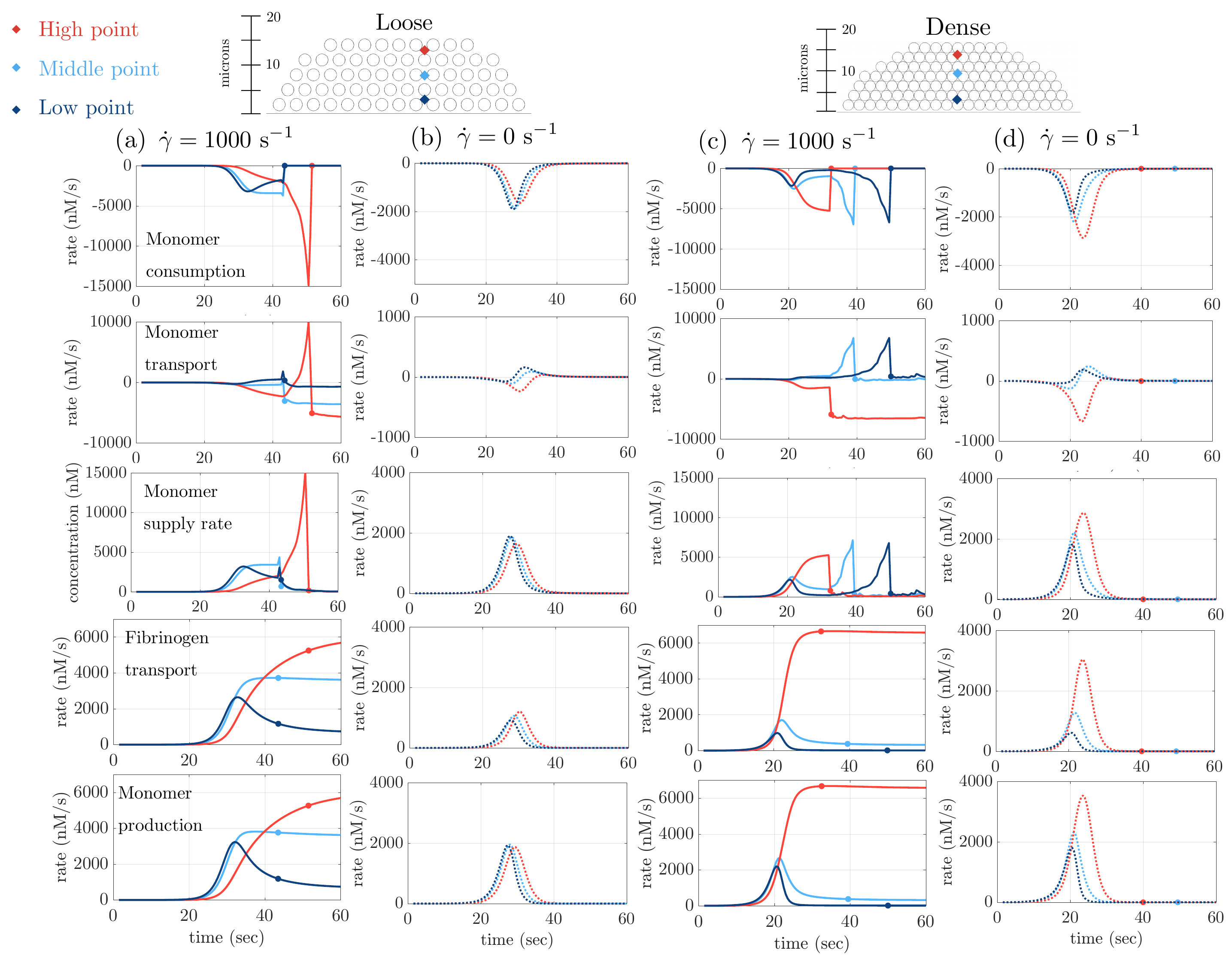}
 \caption{Rates of transport, consumption, and production for the loose platelet plug with (a) flow and (b) without flow  conditions and for the dense platelet plug with (c) flow and (d) without flow. Each color refers to a spatial location within the platelet plug: high point (red), middle point (cyan), and low point (blue). For each simulation, the dot corresponds to the time of gelation. Note that if there is no marker, gelation does not occur before 60 seconds. } 
 \label{S9_fig}
 \end{adjustwidth}
        \end{figure}

\paragraph*{Flow comparison for three spatial points for coagulation species and fibrinogen.}
Here, we show concentrations of coagulation species, \Cref{S7_fig}, and fibrin polymerization species, \Cref{S8_fig} along with the rates of fibrinogen transport, fibrinogen conversion to fibrin monomers, monomer transport, and monomer incorporation into larger oligomers, \Cref{S9_fig}, in the case of flow and no flow. In brief, rates of fibrin monomer supply and consumption and the fibrin species concentrations are smaller for the no flow case for both plug densities.  Coagulation species concentrations for the dense plug are largely the same with and without flow.  For the loose plug, enzymatic species accumulate more  without flow because of a smaller degree of washout.  The difference in the fibrin species concentrations and monomer supply and consumption rates between the flow and no-flow  cases is larger for the loose plug than for the dense one due to very low intraplug velocity in the dense plug. These differences in polymerization dynamics result in differences in the gelation time.  For the loose plug without flow, gelation does not occur at any of the three spatial locations within the first 60 seconds.  For the dense plug  without flow, gelation occurs within 60 seconds at the top and middle point but not at the bottom point while it occurs at all three locations for the dense plug with flow. .

\end{document}